\documentclass[twocolumn]{aastex63}
\usepackage{hyperref}
\usepackage[colorinlistoftodos]{todonotes}
\usepackage{amsmath}
\usepackage{nicefrac}
\usepackage{bm}
\shorttitle{Multi-wavelength constraints on feedback}
\usepackage[encapsulated]{CJK}

\newcommand{\Mpch}{\ensuremath{\mathrm{Mpc}~h^{-1}}}

\newcommand{\Msol}{$M_\odot$}

\newcommand{\eightsigma}{$f_\mathrm{gas}~-8\sigma$}
\newcommand{\foursigma}{$f_\mathrm{gas}~-4\sigma$}
\newcommand{\foursigmajet}{Jet\_$f_\mathrm{gas}~-4\sigma$}

\newcommand{\fgas}{$f_\mathrm{gas}$}

\begin{document}

\title{Joint X-ray, kinetic Sunyaev-Zeldovich, and weak lensing measurements: toward a consensus picture of efficient gas expulsion from groups and clusters}

\author[0000-0002-9337-0902]{Jared C. Siegel}
\altaffiliation{NSF Graduate Research Fellow}
\affiliation{Department of Astrophysical Sciences, Princeton University, 4 Ivy Lane, Princeton, NJ 08544, USA}
\email{siegeljc@princeton.edu}

\author[0000-0002-6445-0559]{Alexandra Amon}
\affiliation{Department of Astrophysical Sciences, Princeton University, 4 Ivy Lane, Princeton, NJ 08544, USA}

\author[0000-0002-1286-483X]{Ian G. McCarthy}
\affiliation{Astrophysics Research Institute, Liverpool John Moores University, Liverpool, L3 5RF, UK}

\author[0000-0002-9870-3331]{Leah Bigwood}
\affiliation{Institute of Astronomy and Kavli Institute for Cosmology, University of Cambridge, Madingley Road, Cambridge, CB3 0HA, UK}
\affiliation{Kavli Institute for Cosmology (KICC), University of Cambridge, Madingley Road, Cambridge CB3 0HA, UK}

\author[0000-0003-1585-997X]{Masaya Yamamoto}
\affiliation{Department of Astrophysical Sciences, Princeton University, 4 Ivy Lane, Princeton, NJ 08544, USA}

\author[0000-0002-7619-5399]{Esra Bulbul}
\affiliation{Max Planck Institute for Extraterrestrial Physics, Giessenbachstrasse 1, 85748 Garching, Germany}

\author[0000-0002-5612-3427]{Jenny E. Greene}
\affiliation{Department of Astrophysical Sciences, Princeton University, 4 Ivy Lane, Princeton, NJ 08544, USA}

\author[0000-0002-4475-3456]{Jamie McCullough}
\affiliation{Department of Astrophysical Sciences, Princeton University, 4 Ivy Lane, Princeton, NJ 08544, USA}

\author[0000-0002-2395-4902]{Matthieu Schaller}
\affiliation{Leiden Observatory, Leiden University, PO Box 9513, 2300 RA Leiden, the Netherlands}

\author[0000-0002-0668-5560]{Joop Schaye}
\affiliation{Leiden Observatory, Leiden University, PO Box 9513, 2300 RA Leiden, the Netherlands}

\begin{abstract}

There is no consensus on how baryon feedback shapes the underlying matter distribution from either simulations or observations.
We confront the uncertain landscape by jointly analyzing new measurements of the gas distribution around groups and clusters---DESI+ACT kinetic Sunyaev-Zel’dovich (kSZ) effect profiles and eROSITA X-ray gas masses---with mean halo masses characterized by galaxy-galaxy lensing. 
Across a wide range of halo masses ($M_{500}=10^{13-14}M_\odot$) and redshifts ($0<z<1$), we find evidence of more efficient gas expulsion beyond several $R_{500}$ than predicted by most state-of-the-art simulations. 
A like-with-like comparison reveals all kSZ and X-ray observations are inconsistent with the fiducial $1$~Gpc$^{3}$  hydrodynamical FLAMINGO simulation, which was calibrated to reproduce pre-eROSITA X-ray gas fractions: eROSITA X-ray gas fractions are $2\times$ lower than the simulation, and the kSZ measurements are combined $>8 \sigma$ discrepant. 
The FLAMINGO simulation variant with the most gas expulsion, and therefore the most suppression of the matter power spectrum relative to a dark matter only simulation, provides a good description of how much gas is expelled and how far it extends;
the enhanced gas depletion is achieved by more powerful but less frequent AGN outbursts.
Joint kSZ, X-ray, and lensing measurements form a consistent picture of gas expulsion beyond several $R_{500}$, implying a more suppressed matter power spectrum than predicted by most recent simulations. 
Complementary observables and next-generation simulations are critical to understanding the physical mechanism behind this extreme gas expulsion and mapping its impact on the large-scale matter distribution.

\end{abstract}


 \section{Introduction}
\label{sec:intro}

Baryon feedback redistributes gas relative to the underlying dark matter distribution. 
Active galactic nuclei (AGN), supernova, and stellar winds all heat and expel gas from halos, while cooling incites halo contraction.
For massive galaxies ($\gtrsim L^\star$), AGN are the dominant feedback mechanism.
AGN feedback was originally motivated by the observed coevolution of supermassive black holes with their host galaxies \citep[e.g.,][]{Silk1998,Springel2005,Kormendy2013, Heckman2014}.
For groups and clusters, energy injection by AGN prevents efficient cooling of the gas reservoir \citep[e.g.,][]{McNamara2007,Conroy2008,McCarthy2011,Fabian2012,Gitti2012,Gaspari2020,Eckert2021}.
On Mpc scales ($k > 0.1~h~\mathrm{Mpc}^{-1}$), there is now mounting evidence that gas expulsion impacts the large-scale matter distribution \citep[][]{bigwood2024,Grandis2024baryons, Kovac25,Dalal2025}.
In this paper, we jointly study three measures of the matter distribution---the kinetic Sunyaev-Zel’dovich (kSZ) effect, X-ray cluster gas mass fractions, and galaxy-galaxy lensing (GGL)---to constrain the impact of feedback as a function of radius, halo mass, and redshift. 

How gas is redistributed by feedback on large scales is a fundamental question in galaxy evolution and a critical ingredient for large-scale structure cosmology. 
Modeling the matter distribution on small scales ($k > 0.1~h~\mathrm{Mpc}^{-1}$) requires knowledge of how baryon feedback suppresses the matter power spectrum relative to a dark matter only universe \citep[e.g.,][]{Chisari2019}. 
By constraining how much gas is depleted from halos and how far it is expelled, observations of the gas distribution place powerful constraints on the suppression of the matter power spectrum. 
Power suppression was first constrained with X-ray observations of the intracluster medium (ICM) in groups and clusters.
The baryon fraction is found to increase as a function of radius and halo mass, returning to the universal baryon fraction at large radii of massive clusters \citep{Sun2009, Bulbul2012, Sanderson2013, Lovisari2015, Eckert2016, Eckert2019}.
By more efficiently expelling gas beyond $R_{500}$, stronger feedback processes are expected to lower the typical gas fraction at a given halo mass. 
Prior studies have inferred moderate power suppression ($\sim$2\% at $k=1~h$~Mpc$^{-1}$) by either calibrating cosmological hydrodynamical simulations to X-ray measurements \citep[e.g.,][]{vanDaalen2011,McCarthy2017,Henden2018,vanDaalen2020,Salcido2023,Schaye2023} or by semi-analytic modeling \citep[e.g.,][]{Debackere2020,Giri_2021,bigwood2024,Grandis2024baryons, Kovac25}.

More recently, evidence has emerged from a range of observables that power suppression is more extreme than most simulations predict:
$\sim$10\% at $k=1~h$~Mpc$^{-1}$, either by inferring the suppression from observations \citep{AmonEfstathiou22, Schneider2022, Preston23,bigwood2024,Dalal2025,Kovac25, Reischke2025} or by comparison to simulations \citep{ Hadzhiyska2024photoz, McCarthy2025,ReidkSZ, Hadzhiyska2025}. 
Stacked measurements of the kSZ effect from DESI$+$ACT have been critical to this work \citep{Hadzhiyska2024photoz, ReidkSZ};
after rapid development over the past decade \citep{Schaan_2016, Battaglia_2017, Schaan2021, Mallaby_Kay_2023}, kSZ measurements now offer powerful constraints on the gas content of groups beyond several $R_{500}$.

The latest X-ray measurements also favor strong power suppression.
The first release of the eROSITA all-sky X-ray survey reports lower average gas fractions than previously observed at a given halo mass \citep{Liu2022, Bulbul2024}, potentially highlighting the impact of selection effects on prior samples \citep{Seppi2022,Marini2024}.
Stacking of eROSITA images on optically selected groups also yields systematically lower gas fractions \citep{Popesso2024}.

To address the uncertain observational landscape, we adopt a multi-probe view of feedback. 
This paper provides the first joint analysis across kSZ, X-ray, and galaxy-galaxy lensing (GGL) to constrain the gas distribution across a wide range of radial scales, halo masses, and redshifts: X-ray gas fractions probe the low redshift cluster mass regime, while kSZ measurements extend to higher redshifts and lower masses. 
We consider gas mass measurements from the first release of the eROSITA all-sky survey \citep{Bulbul2024} and kSZ effect profiles from SDSS$+$ACT \citep{Schaan2021} and DESI$+$ACT \citep{ReidkSZ}, described in Section~\ref{sec:data}, in comparison to the $1$~Gpc$^{3}$ FLAMINGO hydrodynamical simulations \citep{Schaye2023, Kugel2023}, which we summarize in Section~\ref{sec:sim}.
We produce new galaxy-galaxy lensing measurements using overlapping weak lensing data from the Dark Energy Survey, the Kilo-Degree Survey, and the Hyper-Supreme Camera Survey \citep{Gatti_2021, Giblin_2021, Li2022} to measure the mean halo masses of the X-ray and kSZ samples. 
As described in Section~\ref{sec:characterizing_the_samples}, this resolves two key uncertainties in interpreting these samples---the halo mass and the satellite fraction, which are degenerate with the inferred magnitude of gas expulsion---and enables a like-with-like comparison to the simulations \citep{McCarthy2025}. 
We present our simulation comparison in Section~\ref{sec:results} and discuss our findings on the magnitude of matter power suppression in Section~\ref{sec:discussion}.
We conclude in Section~\ref{sec:conclusions}.
In Bigwood et al. in prep. we expand our analysis of the kSZ effect to several additional hydrodynamical simulations.

\section{Observations}\label{sec:data}
\subsection{kSZ effect profiles}\label{sec:kSZ}

CMB photons inverse Compton scatter off the intervening ionized gas around galaxies and clusters.
Due to the bulk motion and velocity dispersion of the gas, these interactions impart a Doppler shift to the CMB photons: the kinetic Sunyaev-Zel’dovich \citep[kSZ,][]{Sunyaev1980} and thermal Sunyaev-Zel’dovich \citep[tSZ,][]{Sunyaev1972} effects, respectively.
The kSZ effect is proportional to the electron number density and the peculiar velocity of the gas, while the tSZ effect depends on the electron pressure (integrated along the line of sight).
By measuring the kSZ effect as a velocity weighted stack, contamination from the cosmic infrared background, which is uncorrelated with the velocity field, cancels on average; the tSZ effect is relatively more sensitive to contamination \citep[e.g.,][]{Liu2025}.

The kSZ effect manifests as a temperature fluctuation in the CMB \citep[e.g.,][]{Schaan2021}:
\begin{align}
    \frac{ \Delta T_\mathrm{kSZ} ({\bm \theta}) }{ T_\mathrm{CMB} } = - \sigma_\mathrm{T} \int n_e ({\bm \theta},z) \frac{ v_{e, \mathrm{r}} ({ \bm \theta}, z) }{ c } e^{-\tau({\bm \theta},z)} \frac{d \chi}{1+z},
\end{align}
where ${\bm \theta}$ is an angular position on the sky, $\sigma_\mathrm{T}$ is the Thomson scattering cross-section, $c$ is the speed of light, $n_e ({\bm \theta},z)$ is the electron density, $v_{e, \mathrm{r}}({\bm \theta},z)$ is the electron velocity along the line of sight, and $\chi$ is the comoving radial distance.
The optical depth to Thompson scattering between the observer and
redshift $z$ is
\begin{equation}
    \tau({\bm \theta},z) = \sigma_\mathrm{T}  \int n_e ({\bm \theta},z) \frac{d \chi}{1+z}.
\end{equation}
The gas of interest is optically thin, therefore, for one galaxy
\begin{align}
    \frac{ \Delta T_\mathrm{kSZ} }{ T_\mathrm{CMB} } \approx - \tau_\mathrm{gal} \frac{ v_{e, \mathrm{r}, \mathrm{gal}}}{ c }.
\end{align}
The kSZ signal must be stacked across many galaxies to achieve a high signal-to-noise ratio measurement \citep{Schaan2021}.  

In this work, we adopt the stacked kSZ effect profiles of \cite{Schaan2021} and \cite{ReidkSZ}, which used spectroscopic redshift tracers from SDSS and DESI, respectively;
we also consider the measurements of \cite{Hadzhiyska2024photoz} in Appendix~\ref{appendix:photo-ksz}, which used photometric redshifts for the velocity reconstruction.
The CMB temperature maps and galaxy samples underlying these measurements are described in Sections~\ref{sec:ksz_BOSS} and \ref{sec:ksz_DESI}, respectively.

The stacked measurements use compensated aperture
photometry (CAP) filters.
For a galaxy located at ${\bm \theta}$ on the sky, CAP filtering measures the cumulative kSZ effect profile as a function of angular separation from the galaxy $\theta_d$:
\begin{align}
    \label{eqn:CAP}
    \mathcal{T}(\theta_d) &= \int d^2\theta \Delta T_\mathrm{kSZ}(\theta)  W_{\theta_d}(\theta),\\
    W_{\theta_d}(\theta) &= \begin{cases}
        1 & \text{if } \theta < \theta_d\\
        -1 & \text{if } \theta_d \leq \theta \leq \sqrt{2}\theta_d\\
        0 & \text{if } \theta > \sqrt{2} \theta_d
    \end{cases},
\end{align}
i.e., the temperature fluctuation within $\theta_d$ of the galaxy is summed and then background subtracted by the temperature fluctuations outside $\theta_d$.

Since the kSZ effect is proportional to the velocity of the gas, simply stacking $\mathcal{T}(\theta_d)$ from many galaxies would result in a null signal; galaxies' peculiar velocities are equally likely to be positive or negative.
The stacked kSZ signal is therefore measured by a combination of velocity and inverse-variance weighting \citep{Schaan2021,ReidkSZ}:
\begin{equation}
    \hat{T}_\mathrm{kSZ}(\theta_d) = - \frac{1}{r_{v, \mathrm{bias}}} \frac{ v^\mathrm{rms}_\mathrm{rec} }{c} \frac{ \sum_i \mathcal{T}_i(\theta_d) ( \frac{v_{i, \mathrm{rec}}}{c}) \sigma^{-2}_i }{ \sum_i ( \frac{v_{i, \mathrm{rec}}}{c})^2 \sigma^{-2}_i },
\end{equation}
where $v_{i, \mathrm{rec}}$ is the estimated line of sight peculiar velocity for the $i$th galaxy, $v^\mathrm{rms}_\mathrm{rec}$ is the standard deviation of the reconstructed peculiar velocities, and $\sigma_i^2$ is the variance in $\mathcal{T}_i(\theta_d)$.
To account for biases in the peculiar velocity reconstruction, the stacked kSZ signal is divided by the correlation between the true and reconstructed velocities in mock catalogs:
\begin{equation}
    \label{eqn:rvbias}
    r_{v, \mathrm{bias}} = \frac{ \langle v_\mathrm{true} v_\mathrm{rec} \rangle }{ v_\mathrm{true}^\mathrm{rms} v_\mathrm{rec}^\mathrm{rms} }.
\end{equation}

\subsubsection{SDSS BOSS: LOWZ and CMASS}
\label{sec:ksz_BOSS}

\cite{Schaan2021} measure the kSZ effect profiles for two galaxy samples (LOWZ and CMASS) from the Baryon Oscillation Spectroscopic Survey (BOSS) DR10 release \citep{Ahn2014} using the DR5 CMB temperature maps of ACT \citep[][]{Fowler2007, Swetz2011,Thornton2016,Henderson2016} and Planck \citep{Planck2020}.
\cite{Schaan2021} consider the 90 and 150~GHz frequency ACT maps.
We adopt the results derived from the 150~GHz temperature map due to the higher signal-to-noise ratio and higher resolution beam ($1.3'$ compared to $1.6'$ for 90~GHz).

The LOWZ (``low redshift") sample includes red galaxies between $0.15 < z < 0.4$. 
The CMASS (``constant mass") sample targets massive galaxies in the redshift range $0.4 < z < 0.8$ \citep{Ahn2012}. 
To avoid contamination by the tSZ signal of massive halos, the few thousand most massive galaxies are omitted, resulting in approximately
$140,000$ and $350,000$ galaxies for LOWZ and CMASS, respectively;
\cite{Schaan2021} reject $M_\mathrm{virial}>10^{14}$~{\Msol} halos using the stellar to halo mass relation of \cite{Kravtsov2018}.
By comparing the true and recovered velocities in mock BOSS catalogs \citep{Manera2013, Manera2015}, \cite{Schaan2021} derive a peculiar velocity bias factor of $r_{\rm v}=0.7$.

\subsubsection{DESI: BGS and LRG}
\label{sec:ksz_DESI}

We consider the \cite{ReidkSZ} kSZ effect profiles for two samples of galaxies from DESI~DR1: the Bright Galaxy Survey \citep[BGS,][]{Hahn2023BGS} and the Luminous Red Galaxy sample \citep[LRG,][]{Zhou2023LRG}.
CMB temperature fluctuations are measured from the ACT~DR6 \citep{Naess2025} and Planck \citep{Planck2020} maps. 
To mitigate tSZ contamination, \cite{ReidkSZ} mask temperature outliers in the ACT map.

The BGS sample is magnitude limited and spans $0 < z < 0.6$.
Following \cite{DESItwopoint2024}, \cite{ReidkSZ} limit the BGS sample to $0.1 < z < 0.4$ and apply a luminosity selection as a function of redshift, to create a sample with an approximately constant number density.
The BGS kSZ stack consists of $96,000$ galaxies. For this sample,
\cite{ReidkSZ} adopt the peculiar velocities of \cite{DESIDR1velocities2025} and set the reconstruction bias to $r_{\rm v} = 0.64$, with an estimated $3\%$ uncertainty \citep{Hadzhiyska2024}.

The LRG sample consists of massive quenched galaxies between $0.4 < z < 1.1$.
\cite{ReidkSZ} divide the LRG sample into four stellar mass bins: $\log_{10} M_*/M_\odot$=$(10.5, 11.2), (11.2, 11.4), (11.4, 11.6),$ and $(11.6, 12.5)$.
The mass selection is performed on stellar masses from \cite{Zhou2023LRG}, which are derived from DESI Legacy Imaging photometry via a random forest algorithm trained on the Stripe 82 Massive Galaxy Catalog \citep{Bundy2015}.
\cite{ReidkSZ} set the peculiar velocity reconstruction bias for the LRGs to $r_v = 0.65$, with an estimated $2\%$ uncertainty \citep{Hadzhiyska2024}.

\cite{ReidkSZ} use the harmonic-space Internal Linear Combination (hILC) CMB temperature map, which combines the multiple ACT~DR6 channels ($90, 150,$ and $220$~GHz) with Planck \citep{Coulton2024};
unlike \cite{Schaan2021}, where the kSZ effect profiles were measured separately for the ACT channels.
Before combining the individual frequency maps, each is convolved with a Gaussian beam ($1.6'$ FWHM).

We also consider the DESI$+$ACT kSZ measurements of \cite{Hadzhiyska2024photoz}, which use the photometric sample of LRG from the DESI Legacy Imaging Survey. \cite{Hadzhiyska2024photoz} adopt the photometric redshifts of \cite{Zhou2023} and use the same ACT$+$Planck CMB temperature map, velocity reconstruction pipeline \citep{Eisenstein2007,White2015}, and stellar mass estimates \citep{Zhou2023LRG} as \cite{ReidkSZ}. \cite{Hadzhiyska2024photoz} measure the kSZ effect profile for four bins of stellar mass using the Main LRG sample: $\log_{10} M_*/M_\odot$=$(11,11.25),(11.25,11.5),(11.5,12),$ and $(12,13.5)$; we omit the highest mass bin because of its small sample size ($<3,000$ galaxies) and potential tSZ contamination. 

In this work, we analyze the spectroscopic measurements for our fiducial study and present a full investigation of the photometric measurements in Appendix~\ref{appendix:photo-ksz}.

\subsection{eROSITA X-ray gas mass fractions}
\label{sec:xray}

The eROSITA satellite \citep[extended ROentgen Survey with an Imaging Telescope Array,][]{Predehl2021}, a soft X-ray telescope on board the Spectrum-Roentgen-Gamma observatory  \citep[SRG,][]{Sunyaev2021}, provides comprehensive X-ray data for a large and well-defined sample of thousands of galaxy groups and clusters in the Western Galactic Hemisphere. 
The first eROSITA All-Sky Survey catalog (eRASS1) includes more than $12,000$ optically confirmed X-ray groups and clusters between the local Universe and $z \approx 1.3$ \citep{Bulbul2024, Kluge2024} selected from the $1,300,000$ total eROSITA X-ray detections \citep{Merloni2024}. 
The eRASS1 groups and clusters are X-ray detected and optically confirmed with the \texttt{eROMaPPer} algorithm
\citep{ Rykoff2014, Rykoff2016,Kluge2024}. 
Photometric redshifts are adopted from \texttt{eROMaPPer} and spectroscopic redshifts, where available. The reported photometric redshift accuracy is $ \delta z /(1 + z) \lesssim 0.005$ for $0.05 < z < 0.9$ \citep{Kluge2024}.

Gas density profiles are derived by fitting a parametric model to the X-ray images using the MBProj2D software, following convolution of surface brightness with the temperature of the ICM through a forward modeling approach \citep{ Sanders2018,Liu2022, Bulbul2024}. 
The gas mass is then calculated by integrating the volume density. 
Total halo masses $M_{500}$ are estimated from a weak lensing calibrated scaling relation between redshift, X-ray count rate, and shear.
The scaling relation is calibrated using $2,533$ clusters in the common area of eROSITA with weak lensing surveys; the calibration clusters are drawn from the higher fidelity ``cosmology" sample of $5,249$ clusters between $0.1<z<0.8$ \citep{Bulbul2024}.
The mass estimates include corrections for miscentering and cluster member contamination, as well as calibration from forward modeling of simulated shear profiles \citep{Grandis2024, Kleinebreil2025, Okabe2025}. 

To validate the reported halo masses, we measure the GGL profile for bins of eROSITA clusters. 
We define four samples of clusters: $\log_{10}M_{500}/M_\odot=(13.3, 14.0)$ and $(14.0, 14.5)$ between $0.05 < z < 0.1$, and $\log_{10}M_{500}/M_\odot=(13.5, 14.0)$ and $(14.0, 14.5)$ between $0.1 < z < 0.2$.
The higher redshift bins coincide with the redshift range used for the eROSITA weak lensing mass calibration, whereas the lower redshift bins explore a redshift and mass range not previously calibrated. 

\begin{figure*}[t!]
\includegraphics[width=\textwidth]{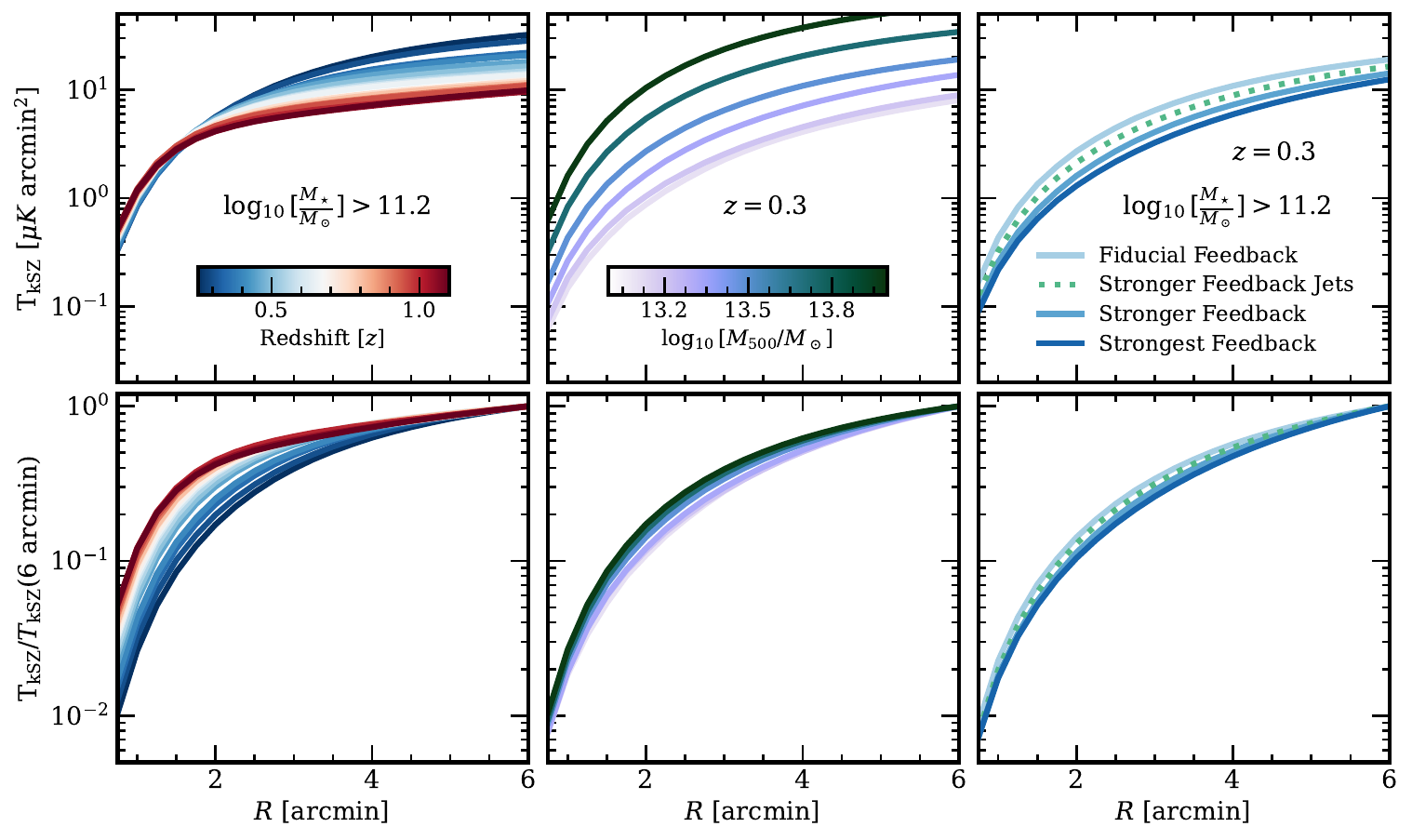}
\caption{Simulated kSZ effect  profiles from FLAMINGO for a range of redshifts (left), halo masses (center), and feedback strengths (right). 
For each kSZ stack, we select simulated halos by a minimum stellar mass cut.
For the left and center panels, we present the kSZ effect from the fiducial FLAMINGO simulation.
On the right, we compare four FLAMINGO simulations of varying feedback strength at a fixed redshift and minimum stellar mass selection.
The bottom row highlights the challenge of inferring feedback strength from the shape of the profile alone: changes in feedback strength are degenerate with the mean redshift and mass of the sample, even when the amplitudes of the kSZ profiles are normalized.
}
\label{fig:ksz_dependence}
\end{figure*}

\section{FLAMINGO simulations}\label{sec:sim}

The FLAMINGO simulation suite includes $16$ hydrodynamical simulations of varied resolution, box size, subgrid modeling, and cosmology \citep{Schaye2023}.
We consider the $1$~Gpc$^3$ intermediate resolution simulations
($m_{\mathrm{gas}} = 1.09\times10^9$ M$_\odot$), with $2 \times 1800^3$ gas and dark matter particles and $1000^3$ neutrino particles.
These simulations adopt the maximum likelihood cosmological parameters of the DES~Y3 `3×2pt + All Ext.' flat $\Lambda$CDM cosmology \citep{DES3x2}.
The simulations successfully reproduce the galaxy stellar mass function, the central black hole--stellar mass relation, the cosmic star formation rate density, and cluster scaling relations \citep{Schaye2023,Braspenning2024}.

For the fiducial simulation, the subgrid physics (e.g., star formation, stellar evolution, radiative cooling, and AGN) were calibrated to reproduce the observed $z=0$ stellar to halo mass relation and the gas fractions of groups and clusters \citep{Kugel2023}.
The gas fraction data was assembled from a collection of pre-eROSITA surveys, including 
\cite{Vikhlinin2006,Maughan2008, Rasmussen2009,
Sun2009,
Pratt2010,
Lin2012,
Lagana2013,
Sanderson2013,
Gonzalez2013,
Lovisari2015,
Hoekstra2015,
Pearson2017,
Mulroy2019,
Lovisari2020,
Akino2022}; with the exception of \cite{Akino2022}, selection effects were not accounted for.
Variants with stronger (weaker) baryon feedback were produced by calibrating to gas fraction measurements shifted down (up) by $ N\sigma$, where $\sigma$ is the observational uncertainty on the mean gas fraction relation and $N \in [2,4,8]$. 
The feedback variants were still calibrated to the observed stellar to halo mass relation.

The strength of baryon feedback is primarily regulated by the AGN subgrid model.
Observationally, AGN are found to couple to the surrounding gas through either the jet or radiative modes \citep{Choi2012};
the jet mode is dominant in the most massive black holes and transfers momentum to the gas via collimated outflows, while the radiative mode deposits thermal energy to neighboring gas \citep[][]{ Heckman2014}.
The fiducial FLAMINGO simulation assumes only radiative AGN feedback, following \cite{Booth2009}.
As a black hole grows,
a fraction of the accreted rest-mass energy heats the neighboring gas;
the fraction of energy converted into heat is $\epsilon_\mathrm{r} \epsilon_\mathrm{f} = 0.015$, where $\epsilon_\mathrm{r} = 0.1$ is the radiative efficiency and $\epsilon_\mathrm{f} = 0.15$ is the fraction of the radiated
energy that heats the gas.
Due to the large masses of gas particles, releasing the heat from the AGN each time step causes numerical overcooling \citep{Dalla2012}.
Instead, the feedback energy is stored internally until it can heat the nearest gas particle by $\Delta T_\mathrm{AGN}$.
Increasing $\Delta T_\mathrm{AGN}$ corresponds to more powerful but less frequent AGN outbursts, resulting in stronger baryon feedback. 
The FLAMINGO suite also includes two simulations with jet mode AGN feedback following \cite{Husko2022}. 
The jet variants were only produced for a limited range of feedback strengths (the fiducial gas fractions and $-4 \sigma$), unlike the radiative mode variants, which extend to $-8 \sigma$. 

In this paper, we study four FLAMINGO simulations: fiducial radiative feedback (L1\_m9), stronger radiative feedback ({\foursigma}), stronger jet feedback ({\foursigmajet}), and strongest radiative feedback ({\eightsigma}).

\section{Characterizing the galaxy samples}
\label{sec:characterizing_the_samples}

Measurements of the stacked kSZ effect and X-ray gas mass fractions are sensitive to the redshift and halo mass of the underlying sample, in addition to feedback strength. 
A careful selection of simulated galaxies is critical to comparing kSZ and X-ray observations with simulations: Figure~\ref{fig:ksz_dependence} (upper panel) shows simulated profiles of the kSZ effect from the FLAMINGO simulations for a range of redshifts (left), halo masses (middle), and feedback strengths (right). 
Even when amplitude changes are accounted for, the changes in the shape of the predicted kSZ signal with redshift and halo mass are degenerate with variations in feedback strength.
As a result, mismatches in the mean halo mass or redshift between the observed and simulated samples can bias the inferred feedback strength (lower panel of Figure~\ref{fig:ksz_dependence}).  
Likewise, the enclosed hot gas mass inferred from X-ray observations is expected to vary with the redshift and halo mass of the sample \citep{Braspenning2024}.
This poses a challenge, because selecting a simulated galaxy sample with the same halo mass as an observed sample is nontrivial.

Following \citet{McCarthy2025}, we ensure that the simulated galaxy samples have the same mean halo mass as the observations using GGL:
for a given observable, we select galaxies from the simulation such that their stacked galaxy–galaxy lensing profile is consistent with the measured lensing profile. 

\subsection{Galaxy-galaxy lensing}
\label{sec:ggl}

\begin{figure*}[t!]
\includegraphics[width=\textwidth]{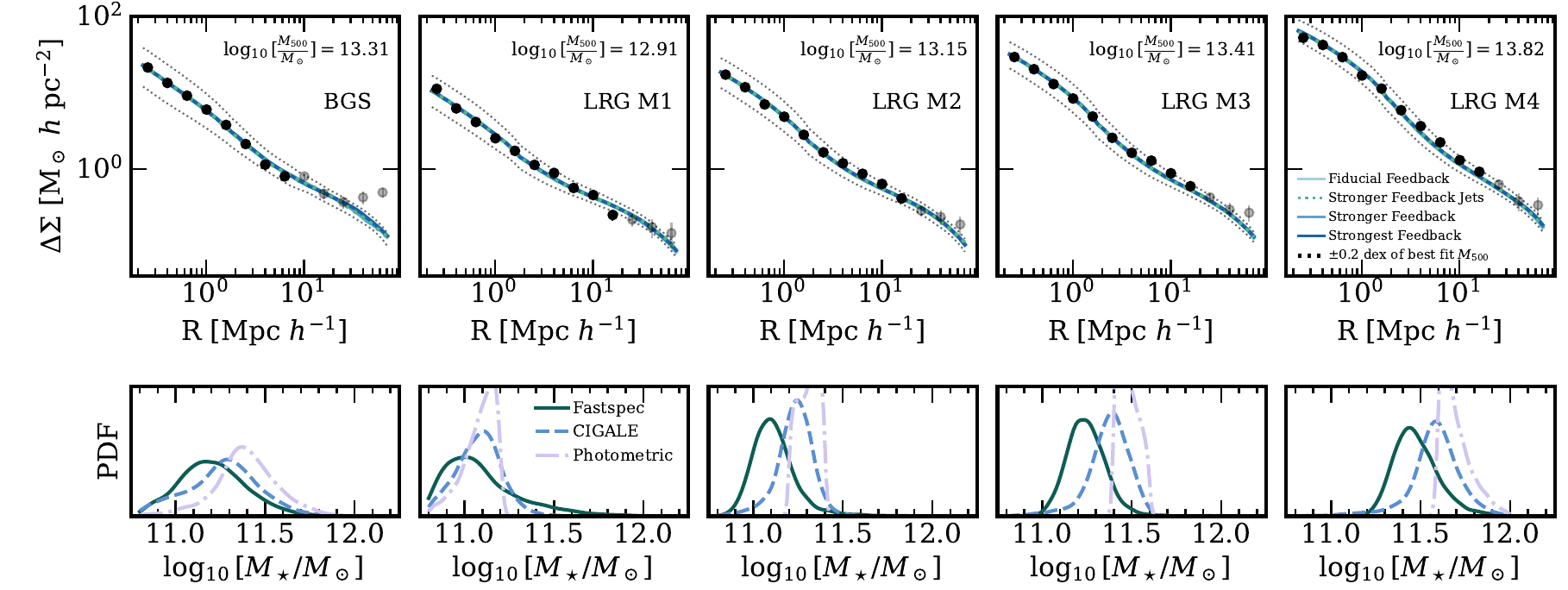}
\caption{
\textit{Top:} The measured GGL profiles (black) of the DESI kSZ samples, alongside the best-fitting FLAMINGO $\Delta \Sigma$ profiles from the four simulations we consider.
The FLAMINGO GGL profiles are calculated by stacking all simulated galaxies above a minimum stellar mass.
At the top of each panel, we report the mean halo mass for the best-fitting selection of simulated galaxies from the fiducial FLAMINGO simulation.
The dotted lines represent the $\Delta \Sigma$ profile for halo masses $\pm 0.2$~dex of the best-fit.
Tangential separations larger than the size of the Jackknife regions (light gray) are omitted from our analysis (Appendix~\ref{app:ggl}). 
\textit{Bottom:} the stellar mass distributions of the DESI samples as measured by \texttt{FastSpecFit}, \texttt{CIGALE}, and the photometric random-forest of \cite{Zhou2023LRG};
\cite{ReidkSZ} defined the LRG mass bins using the photometric estimates.
}
\label{fig:kSZ_wl_calibration}
\end{figure*}

Light from distant background galaxies (sources) is tangentially sheared by foreground structures (lenses), known as galaxy-galaxy lensing (GGL). 
We use GGL to infer the total mass profiles of the SDSS and DESI kSZ samples and the eROSITA X-ray samples, forming the basis of our like-with-like comparisons with the FLAMINGO simulations.

For a foreground lens at redshift $z_{\rm l}$ and comoving distance $\chi_{\rm l}$, the induced tangential shear on a background source at angular separation $\theta$ from the lens is 
\begin{equation}
    \gamma_\mathrm{t}(R = \theta \chi_\mathrm{l}) = \frac{ \Delta \Sigma(R) }{ \Sigma_\mathrm{crit} },
\end{equation}
where $\Delta \Sigma(R) = \Sigma(R) - \bar{\Sigma}(<R) $ is the excess surface density: the difference between the projected surface density at radius $R$ and the average surface density within $R$.
$\Sigma_\mathrm{crit}$ is the comoving critical surface mass density
\begin{equation}
    \Sigma_\mathrm{crit}^{-1} = \frac{4 \pi G}{c^2} \frac{ D_\mathrm{l} D_\mathrm{ls} }{ D_\mathrm{s} } (1+z_\mathrm{l})^2,
\end{equation}
where $D_\mathrm{l}$ and $D_\mathrm{s}$ are the angular diameter distances to the lens and source, respectively, and $D_\mathrm{ls}=D_\mathrm{s}-D_\mathrm{l}$.

We measure GGL through a weighted cross-correlation of the shapes of background sources $\epsilon$ with the positions of foreground lenses.
At a given angular separation $\theta$ between lens and source pairs, the tangential shear estimator is
\begin{equation}
    \langle \gamma_\mathrm{t} (\theta) \rangle = \frac{ \sum_{ \mathrm{ls} } \epsilon_\mathrm{t} w_\mathrm{ls} } { \sum_\mathrm{ls} w_\mathrm{ls} },
\end{equation}
where the summation is over all lens--source pairs separated by an angular separation of $\theta$, $\epsilon_t$ is the tangential ellipticity component of each source, and $w_\mathrm{ls}$ is the combined weight of each lens--source pair. 
Considering an ensemble of source galaxies with a calibrated redshift probability distribution, $n(z_\mathrm{s})$, the average comoving critical surface density is
\begin{equation}
   \overline{\Sigma}_\mathrm{crit}^{-1}(z_\mathrm{l}) = \frac{4\pi G (1+z_{\rm l})^2}{c^2}\int d z_\mathrm{s} n(z_\mathrm{s}) \frac{D_{\rm l}  D_{\rm ls}}{ D_{\rm s}}.
\end{equation}
The average excess surface mass density is then
\begin{equation}
    \label{eqn:raw_esd}
   \Delta \Sigma (R) = \frac{ \sum_\mathrm{ls} \epsilon_\mathrm{t} w_\mathrm{ls}  } { \sum_\mathrm{ls} \overline{\Sigma}_\mathrm{crit}^{-1}(z_\mathrm{l}) w_\mathrm{ls} }.
\end{equation}

\begin{figure*}[t!]
\includegraphics[width=\textwidth]{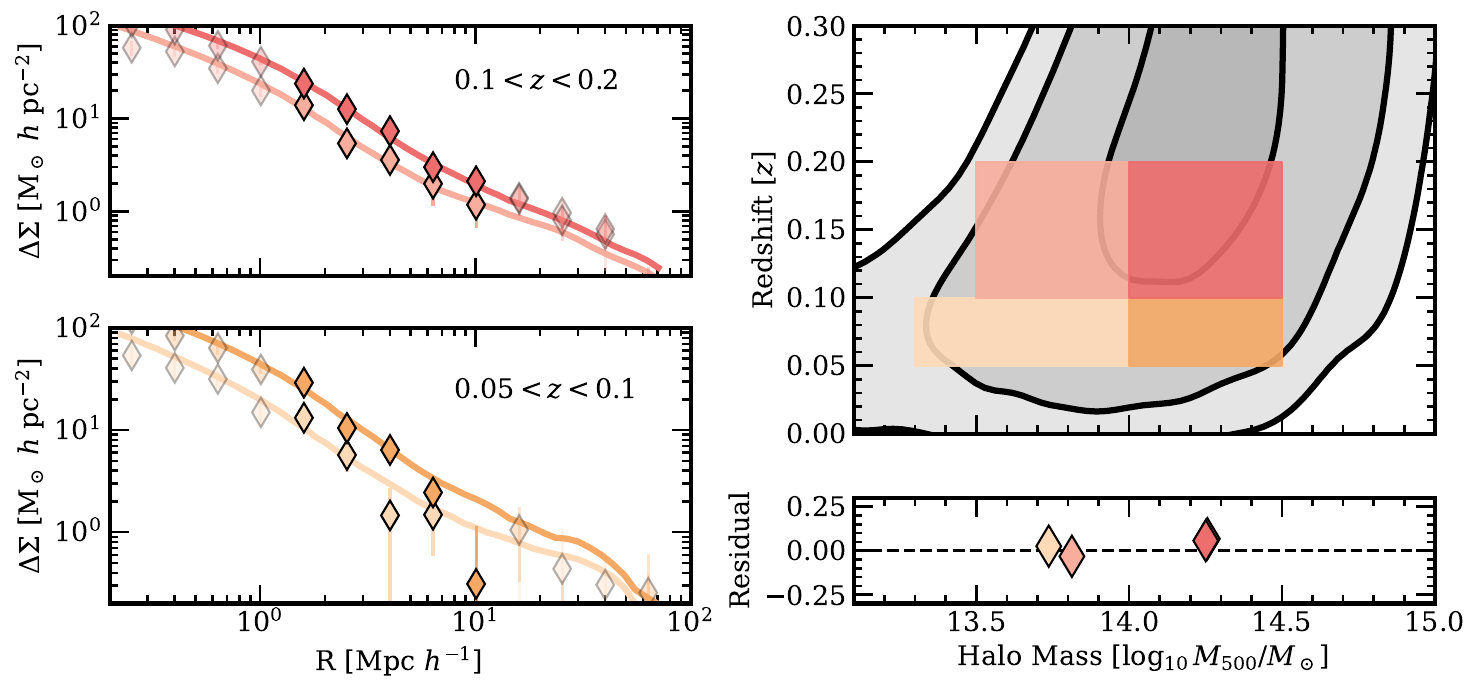}
\caption{
GGL halo mass inference for the eROSITA clusters.
\textit{Left:} measured GGL profiles for $0.05<z<0.1$ (bottom) and $0.1<z<0.2$ (top); for each redshift range, we consider two mass bins, defined in terms of the reported eRASS1 halo masses. 
The transparent points are omitted from the GGL fits (Section~\ref{sec:matching_xray}).
The best-fitting FLAMINGO $\Delta \Sigma$ profiles are presented alongside the observations; for clarity, we only show the fiducial feedback strength simulation.
\textit{Right:} the redshift--halo mass distribution of eRASS1 (gray); contours correspond to the $39, 86,$~and $97$th percentiles.
The definitions of our GGL bins are demarcated by shaded regions.
The lower panel presents the residuals between the GGL derived mean halo masses and the eRASS1 catalog. The halo mass uncertainties are approximately $0.1$ and $0.05$ dex for the low and high mass bins, respectively (Table~\ref{tab:eROSITA_samples}); the errorbars are smaller than the markers. 
}
\label{fig:Xray_wl_calibration}
\end{figure*}

The comoving $\Delta \Sigma (R)$ measurements for the DESI kSZ and eROSITA X-ray samples are presented in Figures~\ref{fig:kSZ_wl_calibration} and \ref{fig:Xray_wl_calibration}, respectively.
For the DESI lenses, we apply the same ACT mask and use the same stellar mass estimates \citep{Zhou2023LRG} as in \cite{ReidkSZ}.
The GGL signals are measured with \texttt{dsigma} \citep{Lange2022} using the DES Y3, KiDS 1000, and HSC Y3 shear catalogs.
The covariance matrix for each measurement is estimated with a leave-one-out Jackknife process  (Appendix~\ref{app:ggl}).
For the SDSS kSZ samples, we adopt the GGL measurements of \cite{Amon2023}.
In Appendix~\ref{app:ggl}, we detail the methodology for computing these measurements, including corrections for known systematics.
We demonstrate the robustness of the signals to survey choice, contamination between the lens and source samples (which determines the impact of intrinsic alignment), and calibration systematics.

\subsection{Like-for-like simulated samples}
\label{sec:selecting_simulated_samples}

Following \citet{McCarthy2025}, we calculate simulated GGL profiles from the FLAMINGO lightcone maps.
The observer's lightcone is approximated as concentric shells in comoving distance, see Appendix~A of \cite{Schaye2023}.
At the redshifts of the kSZ and X-ray observations ($z\lesssim1$), the shells are separated by $\Delta z=0.05$.
For each shell, the quantities of interest (e.g., dark matter mass, hot gas mass, kSZ, etc.) are represented by \texttt{HEALPix} maps \citep{Gorski2005}.
At a given redshift, $z_0$, we calculate the $\Delta \Sigma$ profile for each source in the simulated galaxy sample using the total mass map (gas, dark matter, stars,
black holes, and neutrinos) closest in redshift to $z_0$. 
Per-pixel total masses are converted to comoving surface densities based on the pixel's comoving surface area. 
For computational efficiency, we use a high resolution map ($N_\mathrm{side}=16384$)  when calculating the $\Delta \Sigma$ profile on small scales ($<2~\mathrm{Mpc}~\mathrm{h}^{-1}$) and a lower resolution ($N_\mathrm{side}=2048$) map on larger scales.
The individual $\Delta \Sigma$ profiles are then stacked in comoving radial bins between $0.2 - 75~\mathrm{Mpc}~\mathrm{h}^{-1}$.

For a given selection of simulated galaxies, we calculate the stacked $\Delta \Sigma$ profile for all lightcone shells between the minimum, $z_\mathrm{min}$, and maximum redshift, $z_\mathrm{max}$, of the observed sample. 
The final $\Delta \Sigma$ profile is then obtained by marginalizing over the observed sample's redshift distribution $n(z)$, approximated as a weighted sum from
\begin{equation}
    \label{eqn:noz_marginalization}
    \Delta \Sigma(R) = \int_{ z_\mathrm{min} }^{{ z_\mathrm{max} }} dz \Delta \Sigma_{ z }(R)n(z) .
\end{equation}
This redshift marginalization was not included by \cite{McCarthy2025}, however, it has a minimal impact on the inferred halo mass for the samples we consider.

\subsubsection{SDSS and DESI kSZ effect profiles}
\label{sec:matching_ksz}

For the kSZ effect profiles, replicating the SDSS and DESI selection functions would require careful treatment of the simulated stellar populations, including radiation transfer, dust reddening, and nucleosynthesis. 
Alternatively, the mean halo masses of the samples could be inferred from the galaxies' stellar masses (e.g., abundance matching).
However, systematic uncertainties in stellar population synthesis modeling impart a significant uncertainty in the implied mean halo mass.
In Figure~\ref{fig:kSZ_wl_calibration}, we compare the stellar mass distributions for the DESI samples from three publicly available stellar population synthesis catalogs: i) DESI Legacy Imaging derived masses using a random forest algorithm \citep{Zhou2023LRG}\footnote{\url{https://data.desi.lbl.gov/public/ets/vac/stellar_mass/v1/}}, ii) \texttt{FastSpecFit} modeling of the DESI spectra \citep{Moustakas2023}\footnote{\url{https://data.desi.lbl.gov/doc/releases/dr1/vac/fastspecfit/}}, and iii) \texttt{CIGALE} modeling of the DESI spectra \citep{Siudek2024}\footnote{\url{https://data.desi.lbl.gov/doc/releases/edr/vac/cigale/}}. 
Assuming the stellar-to-halo mass relation of the fiducial FLAMINGO simulation, the different stellar mass estimates correspond to $\gtrsim 0.1$~dex variation in the mean halo mass. 

Instead of relying on the stellar mass measurements, we select a sample of simulated galaxies that reproduces the observed GGL profile for each kSZ stack.
\cite{McCarthy2025} demonstrated that selecting all simulated galaxies above a minimum stellar mass cut successfully reproduces the observed GGL profiles of the SDSS LOWZ and CMASS samples; this selection includes both centrals and satellites.
In the simulations, galaxy stellar mass is defined as the total bound stellar mass within a radius of $50$~kpc, centered on the particle with the minimum gravitational potential. 
For each observed sample, we fit for the minimum stellar mass cut that best-fits the observed GGL profile by $\chi^2$ minimization;
as described above, the simulated $\Delta \Sigma$ profile is marginalized over the redshift distribution of the observed sample.

For each DESI kSZ stack, we present the best-fit FLAMINGO GGL signals alongside the measurements in Figure~\ref{fig:kSZ_wl_calibration}, finding good agreement.
The GGL fitting process is performed independently for the different FLAMINGO simulations; the inferred halo masses are nearly identical between the simulations (Tables~\ref{tab:desi_summary} and \ref{tab:sdss_summary}).
Based on the sizes of the Jackknife regions, we restrict the fits to $<9$ and $20$ comoving~{\Mpch} for the BGS and LRG samples, respectively.
In Appendix~\ref{appendix:sensitivity}, we demonstrate that the inferred halo masses are robust to large variations in the GGL fitting process, such as drawing simulated galaxies from a log-normal distribution on stellar mass; we also verify that our halo mass estimates for SDSS LOWZ and CMASS are consistent with \cite{McCarthy2025}.

\subsubsection{eROSITA X-ray gas masses}
\label{sec:matching_xray}

The eROSITA eRASS1 catalog provides halo mass measurements based on weak lensing calibrated scaling relations between count rate and shear \citep{Ghirardini2024}. 
To validate the halo masses, we measure the GGL profile for four bins of eROSITA clusters, see Table~\ref{tab:eROSITA_samples}. 
The GGL measurements for each X-ray sample are presented in Figure~\ref{fig:Xray_wl_calibration} and are described further in Appendix~\ref{app:ggl}. 

The GGL signals are measured with the DES~Y3 shear catalog because of its significantly larger overlap with the eRASS1 footprint (Figure~\ref{fig:ggl_comparison}).
To mitigate cluster member contamination---cluster members can erroneously be included in the background source galaxy sample due to photometric redshift uncertainties \citep{Hoekstra2012,Gruen2014,Dietrich2019,Varga2019}---we also consider the bespoke ``blue" shear catalog of \cite{McCullough2024}; 
blue shear omits red galaxies, which are more likely to be misidentified cluster members (see Appendix~\ref{app:xray_sensitivity}).
Our analysis of the cluster GGL measurements is restricted to $2-14$ comoving~{\Mpch}; the omission of small scales is designed to suppress cluster member contamination, and the cut on large scales is dictated by the size of the Jackknife regions (Appendix~\ref{app:xray_sensitivity}).

For each sample of clusters, we identify a selection of simulated FLAMINGO halos that best-fits the observed GGL profile.
Our fiducial selection function is a minimum halo mass cut, excluding satellites; the derived halo masses are consistent if simulated clusters are instead drawn from a log-normal distribution in halo mass (Appendix~\ref{appendix:sensitivity}).
The best-fit simulated GGL profiles are presented in Figure~\ref{fig:Xray_wl_calibration}, alongside the measured profiles.
For all four samples, the inferred mean halo mass is within $0.1$~dex of the publicly reported mass.

\section{Results: New kSZ and X-ray data compared to FLAMINGO}\label{sec:results}

We perform like-with-like comparisons between observational probes of the gas distribution (kSZ effect profiles and X-ray hot gas fractions) and the FLAMINGO simulations in Sections~\ref{sec:kSZ_result}~and~\ref{sec:Xray_result}, respectively.

\subsection{kSZ effect profiles}
\label{sec:kSZ_result}

\begin{figure*}[t!]
\includegraphics[width=\textwidth]{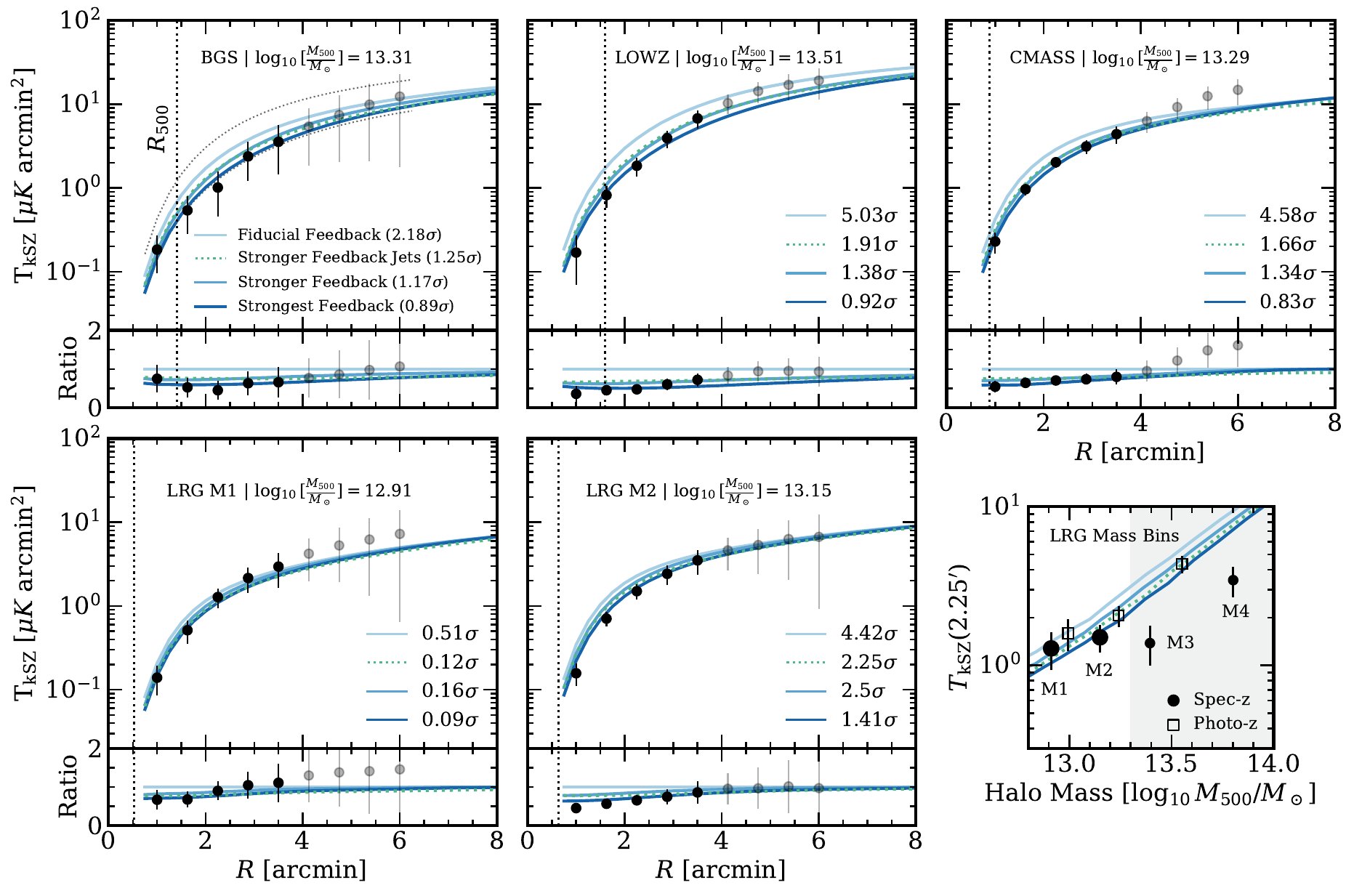}
\caption{
The kSZ measurements (black) compared
with the GGL-selected predictions from the four FLAMINGO simulations we consider. 
We report the number of standard deviations by which each simulation prediction deviates from the observations.
The kSZ effect measurement is highly correlated at larger angular separations, so these bins are shaded light gray to reflect their lower statistical weight.
The lower panels show the ratio of the data and the simulations to the fiducial feedback strength simulation. 
The vertical dotted line demarcates $R_{500}$.
The lower right panel presents the amplitude of the stacked kSZ effect at $\theta=2.25'$ for the spectroscopic \citep{ReidkSZ} and photometric \citep{Hadzhiyska2024photoz} LRG kSZ mass bins ($z=0.75)$ as a function of GGL inferred halo mass, alongside the simulation predictions.
All measurements require more feedback than fiducial FLAMINGO; however, the spectroscopic and photometric measurements require different feedback strengths at $M_{500 }\gtrsim 2\times10^{13}~M_{\odot}$  (shaded region) and are omitted from the primary analysis (see
 Appendix~\ref{appendix:photo-ksz}).}
\label{fig:ksz_result}
\end{figure*}

\begin{figure}[t!]
\includegraphics[width=\columnwidth]{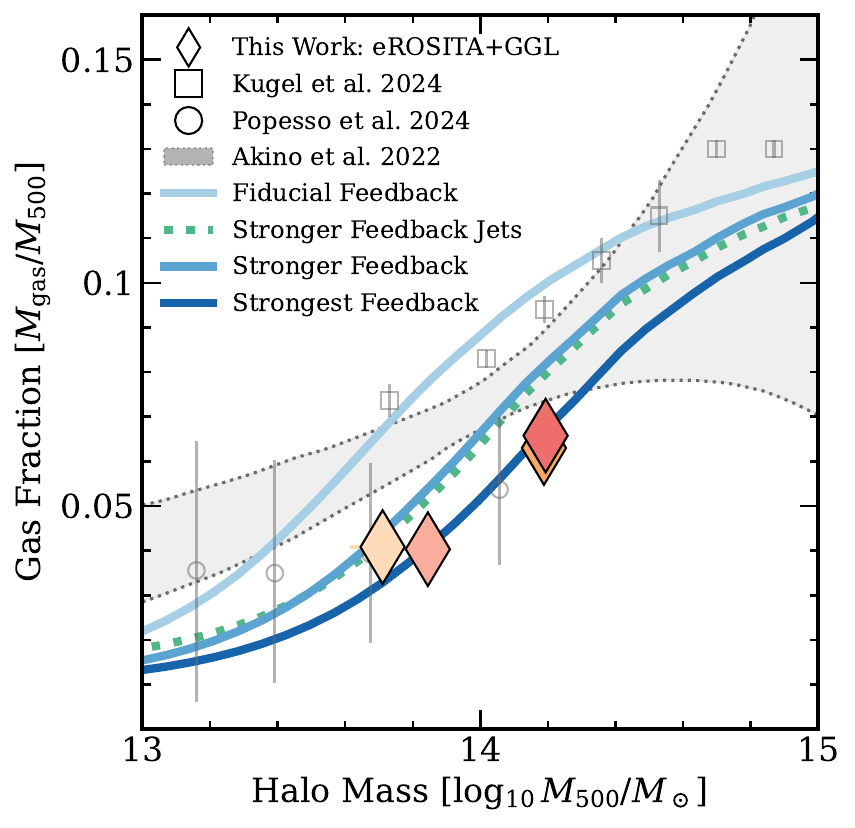}
\caption{
The hot gas fraction $f_\mathrm{gas}$ within $R_{500}$ as a function of halo mass $M_\mathrm{500}$ in groups and clusters.
Our independent halo masses for bins of eROSITA detected clusters are shown as diamonds, with the colors following Figure~\ref{fig:Xray_wl_calibration}.
We report halo mass uncertainties of $<0.1$~dex,  see Table~\ref{tab:eROSITA_samples};
the errorbars are smaller than the markers.
The pre-eROSITA $f_\mathrm{gas}-M_{500}$ relations of \cite{Akino2022} and \cite{Kugel2023} are included for reference, alongside the stacking measurements of \cite{Popesso2024}.
We also present the mean $f_\mathrm{gas}-M_\mathrm{500}$ relation from the four FLAMINGO simulations we consider.
}
\label{fig:gas_fractions}
\end{figure}

To perform a like-with-like comparison, we select a sample of simulated galaxies that reproduces the observed GGL signal for each kSZ measurement (Section~\ref{sec:matching_ksz}).
We measure the corresponding kSZ effect profiles from the simulations following \cite{McCarthy2025}.
At a given redshift, we derive a \texttt{HEALPix} map of the kSZ effect from the Doppler B parameter ($b$) lightcone: $\Delta T_\mathrm{kSZ} = -b T_\mathrm{CMB}.$
To mirror the observations, the kSZ effect maps are convolved with a Gaussian beam: FWHM of $1.3'$ and $1.6'$ for the SDSS and DESI kSZ stacks, respectively.
We then stack the kSZ effect signal for the selected simulated galaxies using the CAP filter (Equation~\ref{eqn:CAP}).
Analogous to the simulated $\Delta \Sigma$ profiles, we marginalize over the redshift distribution of the observed galaxy sample.

The simulated kSZ effect profiles are presented alongside the observations in Figure~\ref{fig:ksz_result}; the vertical lines correspond to $R_{500}$.
For our fiducial analysis, we consider two SDSS kSZ stacks (LOWZ and CMASS) and three DESI kSZ stacks (BGS and two LRG mass bins: M1 and M2).
The kSZ profiles of the fiducial simulation lie significantly above all five measurements.
Since the kSZ effect is proportional to the gas mass, this offset shows that the simulated halos are more gas rich than is favored by the observations.

In the stronger feedback simulations, the AGN more efficiently expel gas beyond the virial radius.
The increased feedback depletes the halos' gas content and lowers the amplitude of the kSZ effect profile.
At lower redshifts ($z<0.6$: BGS, LOWZ, and CMASS), the strongest feedback simulation ({\eightsigma}) closely matches the observations, consistent with \cite{McCarthy2025}.
At $z=0.75$, the two lower mass LRG bins (M1 and M2) are also well described by the {\eightsigma} FLAMINGO simulation. 
Combining all five kSZ measurements, the fiducial FLAMINGO simulation is $>8\sigma$ from the data.
The stronger feedback simulations ({\foursigma} and {\foursigmajet}) provide a better fit but still deviate by approximately $3.5\sigma$.
The strongest feedback simulation reproduces the measurements best ($2 \sigma$ from the data).
The goodness of fits are calculated using the full covariance matrices of the measurements to account for the strong correlations in the outermost bins.

We exclude the highest mass LRG bins (M3 and M4) from our fiducial analysis.
At these masses ($M_{500} > 2 \times 10^{13}$~{\Msol}), the measurements of \cite{ReidkSZ}, which use spectroscopic redshifts, and the measurements of \cite{Hadzhiyska2024photoz}, which use photometric redshifts, imply different feedback strengths; at lower masses, the two methods agree well.
The two highest mass LRG bins (M3 and M4) from the spectroscopic measurements require more gas expulsion ($>4\sigma$ from the data) than the strongest feedback FLAMINGO simulation ({\eightsigma}), but the photometric measurements are well matched by {\eightsigma}  (within $1\sigma$ of the data).\footnote{For the spectroscopic kSZ measurements, we report goodness of fit relative to the simulation predictions using the full measurement covariance matrices. For the photometric measurements, only diagonal uncertainties are readily available. Because the outermost radial bins are highly correlated, we restrict the goodness of fit calculation for the photometric measurements to $\theta<4'$.  }
The measurements from both methods are incompatible with the fiducial FLAMINGO simulation. 
We analyze the photometric measurements and high mass spectroscopic bins in Appendix~\ref{appendix:photo-ksz}.
Although a full investigation is beyond the scope of this work, further study of observational uncertainties and potential systematics in the simulation comparison is warranted.

From our like-with-like comparisons, all observed kSZ effect profiles are discrepant with the fiducial FLAMINGO simulation; the fiducial simulation over predicts the amplitude of the kSZ signal, showing the halos are too gas rich.
Of the FLAMINGO simulations, the observations are best-fit by the strongest feedback variant ({\eightsigma}).

\subsection{X-ray gas fractions}
\label{sec:Xray_result}

The {\fgas}$-M_{500}$ trend is indicative of the strength of feedback and has been used to calibrate cosmological hydrodynamical simulations \citep[e.g.,][]{Kugel2023}.
The eROSITA survey has identified thousands of X-ray groups and clusters between $0.0 \lesssim z \lesssim  1.3$ and $10^{12} \lesssim M_{500} / M_\odot \lesssim 10^{15}$.
The lower mass halos ($\lesssim 10^{14}$~{\Msol}) are particularly informative for studying feedback because the ICM is most depleted in these shallower potential wells.
However, detecting and characterizing low mass clusters is observationally challenging. 
These clusters are naturally fainter and less rich, which limits their detection to lower redshifts and increases the contamination fraction.
In Section~\ref{sec:matching_xray}, we validated the eROSITA halo mass estimates with GGL (Figure~\ref{fig:Xray_wl_calibration}). 

Figure~\ref{fig:gas_fractions} compares the halo masses and gas fractions of our four cluster samples with the FLAMINGO simulations.
We adopt the GGL derived halo masses and the eROSITA reported gas mass measurements. 
The eROSITA gas mass measurements assume $R_{500}$ corresponding to the eROSITA halo mass estimates, which differ (marginally) from our GGL measurements;
the difference in enclosed gas fraction is expected to be small \citep{Velliscig2014,Kugel2023}, especially with our GGL halo masses within $<0.1$~dex of the publicly reported eROSITA masses.
The eROSITA gas fractions agree remarkably well with the strongest feedback FLAMINGO simulation. 
To reconcile our eROSITA samples with the fiducial simulation, the stacked halo masses would need to be $0.4$~dex lower. 
However, the statistical and systematic uncertainties on our $M_{500}$ measurements are below $0.15$~dex.
We discuss potential systematics in the X-ray observations in Section~\ref{sec:xray_disc}.

\section{Discussion}\label{sec:discussion}

\begin{figure*}[t!]
\includegraphics[width=\textwidth]{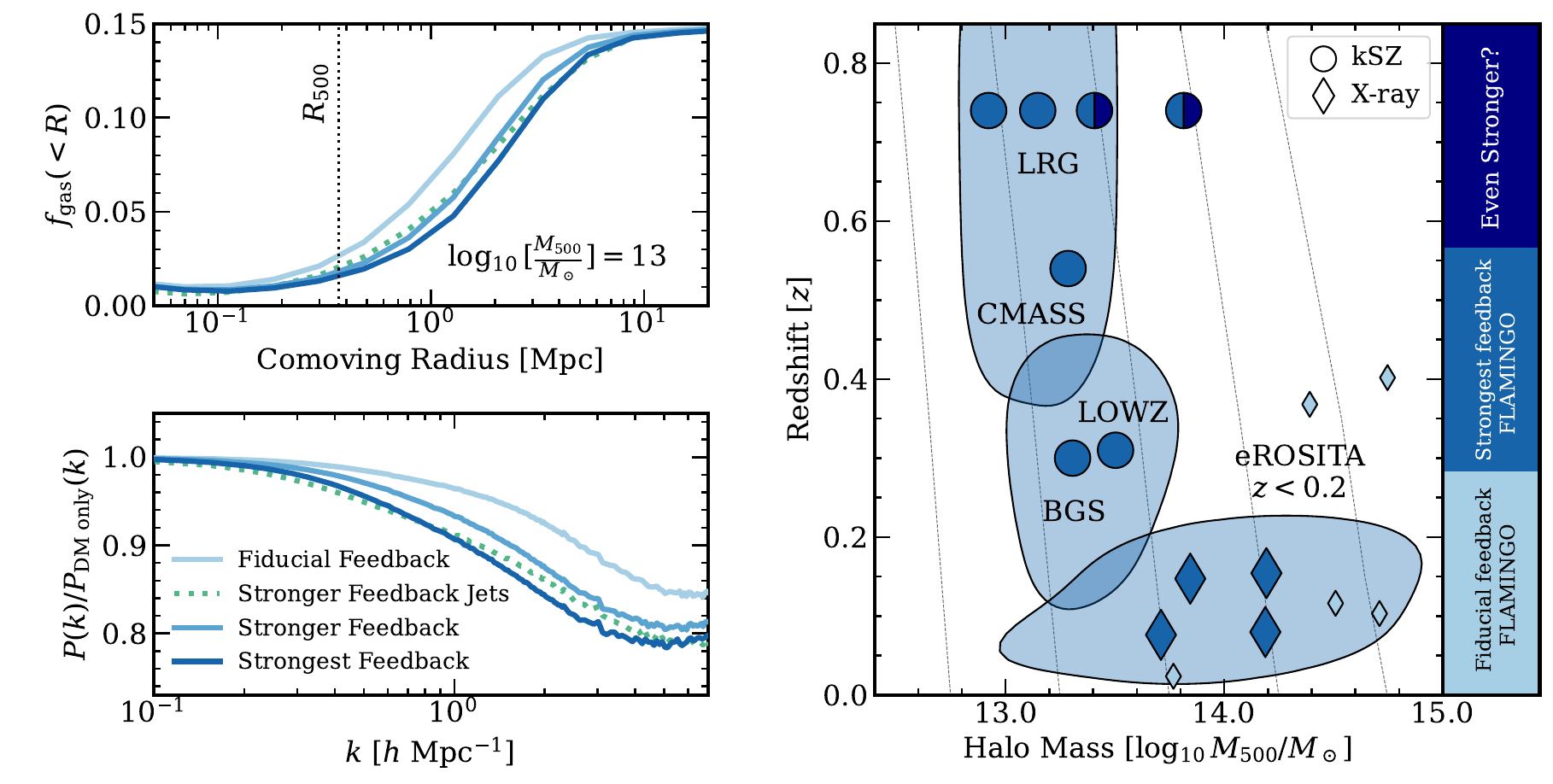}
\caption{
Across a wide range of redshift ($z < 1$) and
halo mass ($\log_{10}M_{500}/M_\odot = 13- 14$), the gas distribution is well described by the strongest feedback FLAMINGO simulation ({\eightsigma}).
\textit{Left:} demonstration of how the different FLAMINGO feedback prescriptions impact the gas distribution. 
For the four simulations we consider, the top panel presents the enclosed gas fraction profile of a $M_{500}=10^{13}~M_\odot$ halo, and the bottom panel shows the matter power spectrum relative to a dark matter only universe.
\textit{Right:} the average redshifts and halo masses of the kSZ stacks and our bins of eROSITA clusters, colored by the preference for fiducial (light blue), strongest (blue), or even stronger feedback (dark blue), in terms of the FLAMINGO suite.
We include the distributions of the DESI BGS and LRG samples, as well as the eROSITA eRASS1 clusters ($z<0.2$); contours correspond to the $86$th percentile.
The mean redshifts and halo masses for a selection of pre-eROSITA X-ray surveys are included in light blue, reflecting their apparent preference for the fiducial simulation \citep{Vikhlinin2006,Maughan2008,Gonzalez2013,Lovisari2015,Eckert2016}.
The dotted lines show the mean growth histories of halos, binned by their redshift zero mass.
}
\label{fig:feedbackmap}
\end{figure*}

We find a consistent picture of strong baryon feedback across halo mass ($10^{13}<M_{500}/M_\odot < 10^{14}$) and redshift ($z<1$), see Figure~\ref{fig:feedbackmap}.
Our joint analysis of kSZ, X-ray, and GGL measurements reveals that the fiducial FLAMINGO simulation does not expel enough gas beyond $R_{500}$ from groups and clusters. 
This is particularly striking because the FLAMINGO matter power spectrum is more suppressed by baryons than other widely used simulations, such as MillenniumTNG \citep{Pakmor2023}, EAGLE \citep{Schaye2015}, and FABLE \citep{Henden2018}.
For every observation, strong feedback provides a better fit to the data than the fiducial FLAMINGO simulation.

At $z<0.6$, all the observations are remarkably consistent with the strongest feedback simulation ({\eightsigma}).
The simulation simultaneously reproduces the eROSITA$+$GGL X-ray gas fractions ($z\sim0.1$) and the kSZ effect profiles of SDSS LOWZ and CMASS and DESI BGS.
At higher redshifts ($z=0.75$), the lower mass (M1 and M2) LRG kSZ effect profiles are also consistent with {\eightsigma}. 
The highest mass LRG bins (M3 and M4) potentially require even stronger feedback, but we caution that the kSZ profiles from the spectroscopic \citep{ReidkSZ} and photometric \citep{Hadzhiyska2024photoz} based measurements imply different feedback strengths at these masses, see Appendix~\ref{appendix:photo-ksz}.

\subsection{On the consistency of X-ray observations}
\label{sec:xray_disc}

The X-ray gas fractions of groups and clusters are a long standing probe of baryon feedback.
eROSITA observations now indicate that groups and clusters are, on average, more gas-depleted than previously found.
In Figure~\ref{fig:gas_fractions}, we present the mean gas fractions for bins of eROSITA detected clusters.
The fiducial FLAMINGO simulation was calibrated to a collection of pre-eROSITA X-ray gas fractions \citep{Kugel2023}, including a selection of surveys without explicit selection effect corrections:
\cite{Vikhlinin2006,Maughan2008, Rasmussen2009,
Sun2009,
Pratt2010,
Lin2012,
Lagana2013,
Sanderson2013,
Gonzalez2013,
Lovisari2015,
Hoekstra2015,
Pearson2017,
Mulroy2019,
Lovisari2020}; 
the mean gas fraction relation from these surveys is presented in Figure~\ref{fig:gas_fractions}.
\cite{Kugel2023} also considered the $f_\mathrm{gas}-M_{500}$ relation of \cite{Akino2022}, shown as a gray band in Figure~\ref{fig:gas_fractions}.
The \cite{Akino2022} model is derived from the HSC--XXL sample, with GGL measured halo masses and corrections for selection effects.
Our GGL calibrated eROSITA bins lie significantly below the \cite{Kugel2023} relation and also fall below the \cite{Akino2022} model.

Selection effects potentially contribute to the apparent discrepancy between the population of eROSITA clusters and the prior gas fraction measurements.
Previous studies often selected X-ray bright nearby group size halos from the ROSAT All-Sky survey \citep{Truemper1982,Voges1999}.
Because X-ray luminosity is proportional to density and gas mass (along with a temperature dependence), X-ray selected samples are biased to concentrated gas rich halos.
The population of eROSITA detected groups and clusters is also shaped by selection effects; 
at $z<0.2$ the completeness of the eRASS1 catalog falls significantly below $10^{14}~M_{\odot}$ \citep{Seppi2022,Popesso2024a,Marini2024}.

To limit the impact of X-ray selection effects, \cite{Popesso2024} measured the stacked gas fractions of optically selected groups using eROSITA images.
The groups were identified from the GAMA survey using the friends-of-friends algorithm \citep{Robotham2011,Driver2022}. 
While \cite{Popesso2024} mitigates the effects of X-ray selection,  optical group finders can be contaminated by $>30\%$ \citep{Seppi2025}.
The measured $f_\mathrm{gas}-M_{500}$ relation from \cite{Popesso2024} is presented in Figure~\ref{fig:gas_fractions}, alongside our GGL calibrated bins.
Despite the differences in selection, the results of \cite{Popesso2024} are consistent with our GGL calibrated eROSITA gas fractions within $1\sigma$. 
An in-depth study of the entropy of eROSITA-selected galaxy groups, accounting for the selection function, also indicates that many cosmological hydrodynamic simulations (MillenniumTNG, Magneticum, OWLS) exhibit milder feedback than the observations require, consistent with our findings on the gas mass fraction \citep{Bahar2024}.

Halo mass estimation is another source of systematic bias.
Prior studies often derived total halo masses assuming hydrostatic equilibrium.
Contributions from non-thermal pressure can bias hydrostatic masses low by approximately $30\%$ \citep{Hoekstra2015,Eckert2016,Kugel2023,MunozEcheverria2024}; 
this bias is dependent on cluster mass \citep{Braspenning2025}.
The eRASS1 halo masses are estimated from a weak lensing calibrated scaling relation \citep{Ghirardini2024}.
While the scaling relation is limited by the calibration data, our GGL validates the eROSITA reported masses in the previously uncertain low mass regime. 

X-ray flux calibration and the assumed ICM metallicity are potential additional sources of minor bias. 
For instance, \cite{Bulbul2024} found that the eROSITA derived luminosities are approximately $15\%$ lower than previously inferred with \textit{Chandra} for a sample of SPT-selected galaxy clusters; this offset corresponds to gas mass estimates skewed lower by roughly $7\%$ (at fixed temperature, X-ray luminosity is proportional to gas density squared); our results are insensitive to this offset but further study is warranted.
The impact of metallicity assumptions is also expected to be small, contributing less than a few percent to the uncertainty \citep{Liu2022}.

Forward modeling of X-ray and optical selection effects will be critical in reducing systematic uncertainties in the $f_\mathrm{gas}-M_{500}$ relation.
However, a consensus picture of strong feedback is already emerging. 
Optical \citep{Popesso2024} and X-ray \citep{Bulbul2024} selected samples are consistent at $10^{14}~M_{\odot}$ and require more gas depletion than previously believed. 
A comprehensive study of gas mass fractions in eROSITA selected galaxy groups, accounting for the relevant selection effects, is warranted to obtain a complete picture of feedback across a wider mass range (Ding et al. in prep).

\subsection{Implications for feedback models}

Our analysis of kSZ, X-ray, and GGL measurements reveals that more gas is expelled beyond $R_{500}$ across redshift ($z<1$) and mass ($10^{13}<M_{500} /M_\odot < 10^{14}$) than predicted by the fiducial FLAMINGO simulation, even though the gas fractions in the fiducial simulation are actually lower than other cosmological simulations, including MillenniumTNG \citep{Pakmor2023}, EAGLE \citep{Schaye2015}, and FABLE \citep{Henden2018}.
Of the FLAMINGO simulations, the strongest feedback variant ({\eightsigma}) reproduces the kSZ and X-ray observations best.
The landscape of observations that we consider is summarized in Figure~\ref{fig:feedbackmap}:
the eROSITA X-ray gas fractions and SDSS/DESI$+$ACT kSZ samples are shown as a function of redshift and halo mass, alongside a selection of pre-eROSITA X-ray data \citep{Vikhlinin2006,Maughan2008,Gonzalez2013,Lovisari2015,Eckert2016}.
Our results present a consistent picture of strong baryon feedback and suggest that baryons impact the matter power spectrum more than predicted by recent simulations.

Evidence of comparatively strong feedback is mounting \citep{AmonEfstathiou22,Schneider2022,Preston23, bigwood2024,LaPosta2024,McCarthy2025, Pandey25, Dalal2025,Hadzhiyska2025, Kovac25, Reischke2025}. 
Within the context of the FLAMINGO simulations, the strongest feedback model ({\eightsigma}) produces the best match to the kSZ effect and gas fraction measurements examined here.  
However, it should be noted that \cite{Braspenning2024} found that this model deviates from the pre-eROSITA observed cluster scaling relations and the radial profiles of massive clusters ($M_{500}>10^{14.5}~M_\odot$).  
The more moderate feedback variants (e.g., $f_\mathrm{gas}~-4\sigma$) are still consistent with the cluster measurements but do not reproduce the kSZ measurements as well.
It will be interesting to explore whether variations in the implemented feedback modeling can successfully reproduce these observations simultaneously and potentially make distinct predictions for other observables.  
For example, some recent studies have indicated that the properties of clusters are also sensitive to the \textit{mode} of AGN outbursts (e.g., radiative versus jet; see \citealt{Braspenning2024,Bigwood2025XFABLE}). 
As we explore variations in the modeling, we should also investigate possible remaining biases in the observations.  
This necessitates forward modeling the data from the simulations as faithfully as possible, including kSZ extraction from realistic synthetic CMB maps and applying the same velocity reconstruction methods as the observations, as well as selecting simulated groups and clusters with the same method used for the eROSITA X-ray data.  
This is the subject of ongoing work.

\section{Conclusions}
\label{sec:conclusions}

The efficiency and extent of gas expulsion beyond $R_{500}$ and its impact on the large scale matter distribution is unclear. 
To confront this uncertain landscape, a multi-probe view of the gas content of groups and clusters is required. 
The aim of this work is to jointly constrain the gas distribution with kSZ effect profiles, X-ray gas mass fractions, and galaxy-galaxy lensing. We benchmark the measurements against the suite of $1$~Gpc$^{3}$ FLAMINGO simulations \citep{Schaye2023}. 
The main results of this study are: 
\begin{itemize}

    \item Our galaxy-galaxy lensing provides precise mean halo mass constraints for the eROSITA X-ray gas mass measurements \citep{Bulbul2024} and the SDSS/DESI+ACT kSZ effect profiles \citep{Schaan2021, ReidkSZ}. The X-ray and kSZ measurements together span a wide range of halo masses ($M_{\rm 500}=10^{13-14}M_\odot$), redshifts ($0<z<1$), and radial scales ($<2-3~R_{500}$). 
    By jointly analyzing the kSZ and X-ray measurements with GGL, we address the leading uncertainties in constraining the gas distribution: i)  the strength of feedback inferred from the measurements is degenerate with halo mass and satellite fraction and ii) kSZ and X-ray measurements  on their own span limited ranges of halo mass and redshift. 
    We select simulated galaxy samples that reproduce the measured GGL signal to ensure that the halo mass and satellite fraction of the simulated and observed samples match.
    We demonstrate that this calibration is insensitive to the assumed feedback strength in the simulation, as well as changes in the lensing (e.g., choice of imaging survey and selection of background sources); see Appendix~\ref{appendix:sensitivity}. 

    \item The fiducial FLAMINGO simulation is significantly disfavored by the kSZ effect profiles ($>8\sigma$ combined) and the new eROSITA X-ray gas fractions.
    For groups and low mass clusters, FLAMINGO halos are too gas rich within $R_{500}$.
    The fiducial simulation was calibrated on the pre-eROSITA gas fractions of low-redshift galaxy groups and clusters \citep{Kugel2023}, which were potentially biased towards gas rich halos.
    \item The FLAMINGO simulation with the strongest baryon feedback provides a good match to the SDSS/DESI$+$ACT kSZ and eROSITA X-ray measurements.
    This simulation was calibrated to pre-eROSITA gas fractions systematically shifted down, and the strong feedback is achieved by more powerful but less frequent AGN outbursts.
    The success of the strongest feedback simulation (referred to as {\eightsigma}) in describing the gas distribution to several $R_{500}$ across group and cluster masses and between $0<z<1$ forms strong indirect evidence for extreme suppression of the matter power spectrum ($\sim10\%$ at $k=1~h~$Mpc$^{-1}$), independent of the detailed physics of feedback.
\end{itemize}

\textbf{Outlook:} 
Further exploration of feedback implementations is warranted.
The strongest feedback FLAMINGO simulation (\eightsigma) provides an excellent description of where the gas is distributed, while moderate feedback FLAMINGO variants (e.g., $f_\mathrm{gas}~-4\sigma$) are more consistent with observed cluster scaling relations and radial profiles \citep{Braspenning2024} but less compatible with the kSZ measurements.
Combining the five kSZ measurements included in our fiducial analysis, the strongest feedback simulation is $2\sigma$ from the data, while the more moderate simulations ({\foursigma} and {\foursigmajet}) are $3.5\sigma$ from the data.
Since the properties of clusters are also sensitive to the mode of AGN feedback (e.g., radiative versus jet; see \citealt{Braspenning2024,Bigwood2025XFABLE}),
it will be informative to explore whether variations in the subgrid AGN feedback modeling or other mechanisms of gas expulsion \citep[e.g., cosmic ray transport,][]{Quataert2025} can more successfully reproduce all observations simultaneously. 
We expand our analysis of the kSZ effect to several additional hydrodynamical simulations in Bigwood et al. in prep.

In this work, we used GGL to resolve uncertainties in interpreting the kSZ measurements related to the mean halo mass and satellite fraction; however, there are still remaining biases to be explored.
In particular, our analysis omitted the two high mass DESI LRG kSZ bins (M3 and M4), because the kSZ stacks measured from spectroscopic and photometric redshifts imply different feedback strengths at these masses (Appendix~\ref{appendix:photo-ksz}). 
The photometric measurements \citep{Hadzhiyska2024photoz} are well described by the strongest feedback variant ({\eightsigma}), while the spectroscopic measurements \citep{ReidkSZ} appear to require even stronger gas expulsion. 
This warrants further investigation into the measurements that is beyond the scope of this work. 
To more fully account for systematics in future like-with-like simulation comparisons, the simulated peculiar velocities should be reconstructed using the same method as the observations; in this work, we use the true peculiar velocities from FLAMINGO and accounted for inaccuracies in the observed velocities using the correction factors of \cite{Schaan2021} and \cite{Hadzhiyska2024}.

Forward modeling of the X-ray gas fractions is also a necessary step.
X-ray detected samples are incomplete at lower masses ($\lesssim 10^{14}~M_\odot$) and potentially biased toward gas rich halos \citep{Seppi2022,Marini2024}.
In this work, we did not apply an X-ray selection function to the simulated halos and instead limited our comparison to massive halos ($\gtrsim 10^{14}~M_\odot$) from the eROSITA sample, which is more complete than prior X-ray selected surveys.
We demonstrated that our sample of eROSITA detected clusters is consistent with the gas fractions of optically selected halos \citep{Popesso2024}.
Forward modeling the emission from simulated halos is also informative for investigating systematics in the measurement of gas mass from X-ray images.

Accurately constraining the suppression of the matter power spectrum due to baryon feedback is vital to extracting cosmological information from the non-linear regime.
From a consistent analysis of kSZ, X-ray, and GGL measurements, we report a consensus picture of strong gas expulsion: across group and cluster masses and between $0<z<1$, the gas distribution is well described by the strongest feedback FLAMINGO simulation ({\eightsigma}) out to several $R_{500}$.
Independent of the mechanism that redistributes gas on such large radial scales, the agreement between the strongest feedback simulation and the data forms a compelling picture of significant suppression of the matter power spectrum: $\sim10\%$ at $k=1~h~$Mpc$^{-1}$.
This level of suppression is stronger than most state-of-the-art hydrodynamical simulations and has implications for current weak lensing studies and next generation surveys \citep[e.g., Vera Rubin, Euclid, and Roman,][]{Ivezic2019,Euclid2025,Akeson2019}. Future survey data from Simons Observatory and NewAthena will provide the next decisive tests \citep{Nandra2013}.

\acknowledgments

We thank Bernardita Ried Guachalla for helpful feedback and making the data available to us. We also thank Sven Heydenreich for helpful guidance in using DESI data and George Efstathiou and Simone Ferraro for useful feedback on this manuscript.
JS acknowledges support by the National Science Foundation Graduate Research Fellowship Program under Grant DGE-2039656. E. Bulbul acknowledges financial support from the European Research Council (ERC) Consolidator Grant under the European Union’s Horizon 2020 research and innovation program (grant agreement CoG DarkQuest No 101002585). 
Any opinions, findings, and conclusions or recommendations expressed in this material are those of the author(s) and do not necessarily reflect the views of the National Science Foundation. This work was supported by the Science and Technology Facilities Council (grant number ST/Y002733/1).
This work used the DiRAC@Durham facility managed by the Institute for Computational Cosmology on behalf of the Science and Technology Facilities Council (STFC) Distributed Research Utilizing Advanced Computing (DiRAC) High Performance Computing Facility (\url{www.dirac.ac.uk}). 
The equipment was funded by BEIS capital funding via STFC capital grants
ST/K00042X/1, ST/P002293/1, ST/R002371/1, and ST/S002502/1,
Durham University and STFC operations grant ST/R000832/1.
DiRAC is part of the National e-Infrastructure.

The authors are honored to be permitted to conduct scientific research on I'oligam Du'ag (Kitt Peak), a mountain with particular significance to the Tohono O’odham Nation.

\textit{KiDS-1000: }Based on observations made with ESO Telescopes at the La Silla Paranal Observatory under programme IDs 177.A-3016, 177.A-3017, 177.A-3018 and 179.A-2004, and on data products produced by the KiDS consortium. The KiDS production team acknowledges support from: Deutsche Forschungsgemeinschaft, ERC, NOVA and NWO-M grants; Target; the University of Padova, and the University Federico II (Naples).

\textit{DES Y3: }This project used public archival data from the Dark Energy Survey (DES). Funding for the DES Projects has been provided by the U.S. Department of Energy, the U.S. National Science Foundation, the Ministry of Science and Education of Spain, the Science and Technology FacilitiesCouncil of the United Kingdom, the Higher Education Funding Council for England, the National Center for Supercomputing Applications at the University of Illinois at Urbana-Champaign, the Kavli Institute of Cosmological Physics at the University of Chicago, the Center for Cosmology and Astro-Particle Physics at the Ohio State University, the Mitchell Institute for Fundamental Physics and Astronomy at Texas A\&M University, Financiadora de Estudos e Projetos, Funda{\c c}{\~a}o Carlos Chagas Filho de Amparo {\`a} Pesquisa do Estado do Rio de Janeiro, Conselho Nacional de Desenvolvimento Cient{\'i}fico e Tecnol{\'o}gico and the Minist{\'e}rio da Ci{\^e}ncia, Tecnologia e Inova{\c c}{\~a}o, the Deutsche Forschungsgemeinschaft, and the Collaborating Institutions in the Dark Energy Survey.
The Collaborating Institutions are Argonne National Laboratory, the University of California at Santa Cruz, the University of Cambridge, Centro de Investigaciones Energ{\'e}ticas, Medioambientales y Tecnol{\'o}gicas-Madrid, the University of Chicago, University College London, the DES-Brazil Consortium, the University of Edinburgh, the Eidgen{\"o}ssische Technische Hochschule (ETH) Z{\"u}rich,  Fermi National Accelerator Laboratory, the University of Illinois at Urbana-Champaign, the Institut de Ci{\`e}ncies de l'Espai (IEEC/CSIC), the Institut de F{\'i}sica d'Altes Energies, Lawrence Berkeley National Laboratory, the Ludwig-Maximilians Universit{\"a}t M{\"u}nchen and the associated Excellence Cluster Universe, the University of Michigan, the National Optical Astronomy Observatory, the University of Nottingham, The Ohio State University, the OzDES Membership Consortium, the University of Pennsylvania, the University of Portsmouth, SLAC National Accelerator Laboratory, Stanford University, the University of Sussex, and Texas A\&M University.
Based in part on observations at Cerro Tololo Inter-American Observatory, National Optical Astronomy Observatory, which is operated by the Association of Universities for Research in Astronomy (AURA) under a cooperative agreement with the National Science Foundation.

\newpage
\appendix

\section{Galaxy Galaxy Lensing}
\label{app:ggl}

\begin{figure}[t!]
\includegraphics[width=\textwidth]{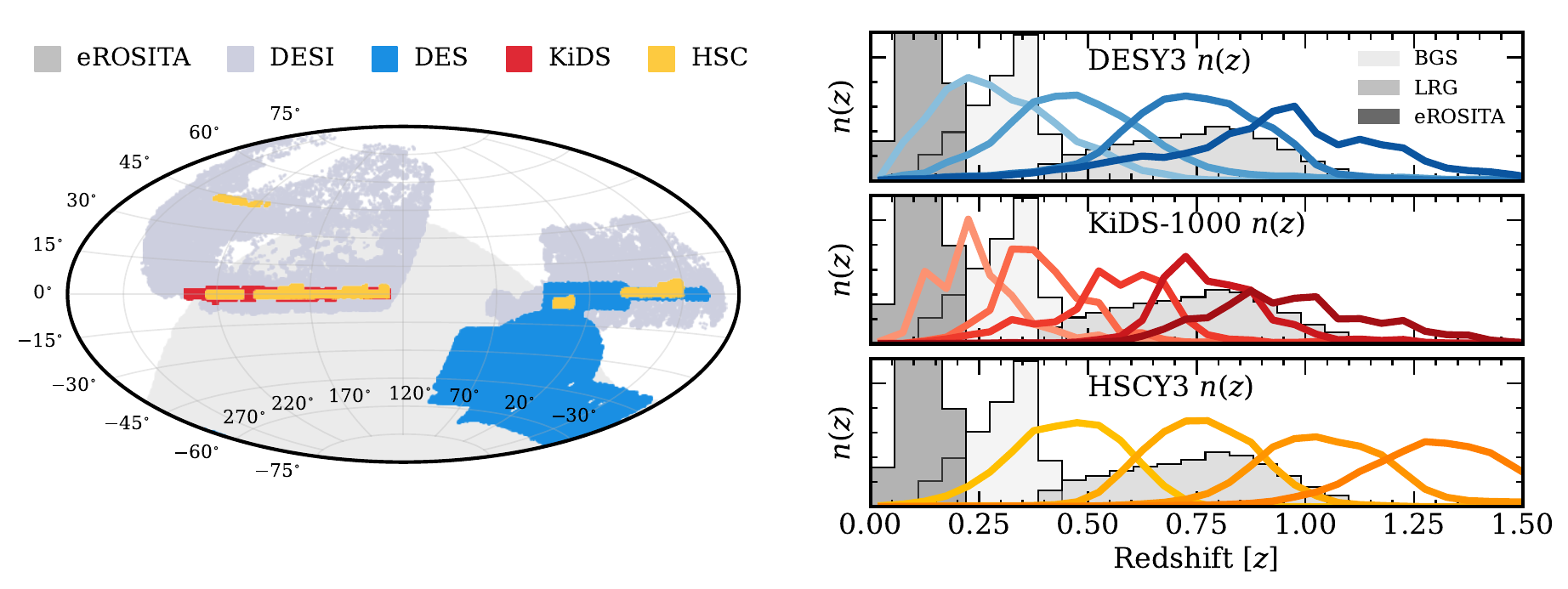}
\includegraphics[width=\textwidth]{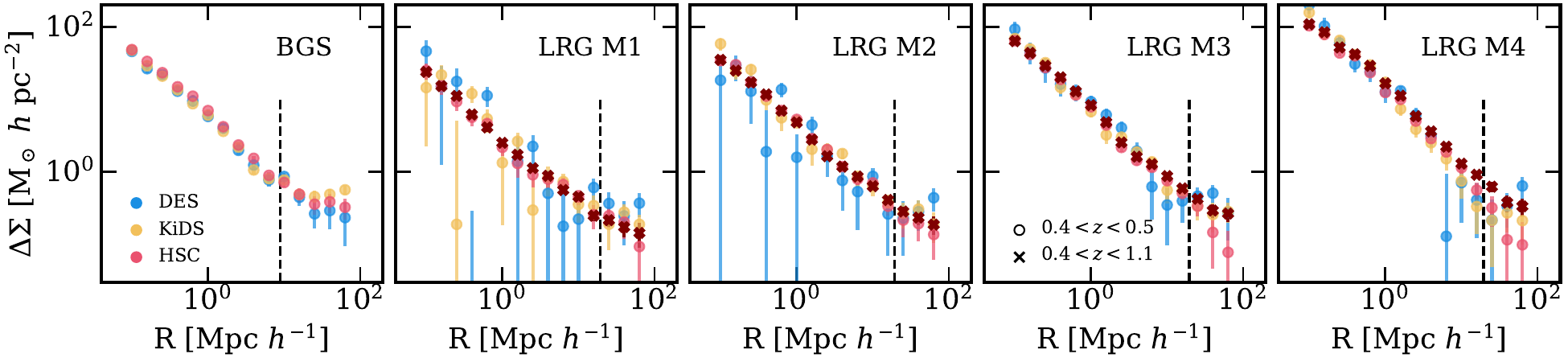}
\caption{
\textit{Upper left:} the footprint of the DESI DR1 data (the BGS and LRG samples are shown in light blue), eRASS1 (gray), and the three imaging surveys: DES (blue), KiDS (red), and HSC (yellow).
\textit{Upper right:} the redshift distributions $n(z)$ of the lenses for our galaxy-galaxy-lensing: eROSITA eRASS1 groups and clusters and DESI~DR1 BGS and LRG.
Each row presents the $n(z)$ of a different lensing survey's tomographic bins: DES~Y3, KiDS~1000, and HSC~Y3.
For measuring the GGL signal, we only consider source tomographic bins sufficiently behind the lens ($\bar{z}_{\mathrm{bin}~i}>z_l+0.1$, where $\bar{z}_{\mathrm{bin}~i}$ is the average redshift of the $i$th tomographic bin and $z_l$ is the lens redshift). 
\textit{Bottom:} the excess surface density measurements for the DESI lenses using the different lensing surveys.
Each column presents a different lens sample.
For the LRG samples, only HSC has a tomographic bin at sufficiently high redshift to measure the GGL signal for the full redshift range ($z=0.4-1.1$); 
we therefore present the GGL measurement for a selection of low redshift lenses ($z=0.4-0.5$) for DES, KiDS, and HSC, alongside the full redshift range measurement from HSC. 
The vertical line shows the size of the Jackknife patches.
}
\label{fig:ggl_comparison}
\end{figure}

In this paper, we measure stacked GGL profiles using three shear catalogs: DES~Y3, HSC~Y3, and KiDS~1000.
The on sky footprints of the lensing surveys are presented in Figure~\ref{fig:ggl_comparison}, alongside DESI~DR1 and eRASS1.
The tomographic redshift distributions, as reported by each survey are also shown in Figure~\ref{fig:ggl_comparison}.
The lensing catalogs are described in Appendix~\ref{app:sources}.
We describe corrections to the GGL estimator in Appendix~\ref{app:lensing_corrections}.

We closely follow the methodology described in \citet{Lange2024} and \citet{Heydenreich2025} for our DESI galaxy-galaxy lensing measurements. For the BGS lenses, we combine the GGL measurements across  DES~Y3, HSC~Y3, and KiDS~1000 by an inverse-variance weighted average.
The calibrated measurements from the individual surveys are statistically consistent within $2 \sigma$ (Figure~\ref{fig:ggl_comparison}).
For the LRG lenses, only HSC has sufficiently high redshift tomographic bins to measure the GGL signal over the full redshift range $z=0.4-1.1$.
To gauge whether the lensing surveys are consistent, we measure the GGL signal for low redshift LRG lenses ($z=0.4-0.5$) with DES, HSC, and KiDS; we again find that the measurements are statistically consistent within $2 \sigma$ (Figure~\ref{fig:ggl_comparison}).
For our analysis of the LRG lenses, we adopt the HSC GGL measurement using the full redshift range $z=0.4-1.1$.
The consistency between the lensing surveys was previously established by \citet{Heydenreich2025}  \citep[also see][]{Amon2023}, which found that the GGL signals from the same data we consider are consistent after correcting the mean redshifts of the HSC tomographic bins; we include the HSC correction following \cite{Li2023} and \cite{Heydenreich2025}.

For the eROSITA lenses, we rely on the DES~Y3 shear catalog, because of its significant overlap with the eROSITA footprint (Figure~\ref{fig:ggl_comparison}).

\subsection{Weak lensing catalogs}
\label{app:sources}

The Dark Energy Survey (DES) imaged over $5000~\mathrm{deg}^2$ of the Southern sky ($grizY$) with the Blanco $4$~meter telescope at Cerro Telolo Inter-American Observatory \citep[]{DES2005}.
We use the DES~Y3 shear catalog,\footnote{\url{https://des.ncsa.illinois.edu/releases/y3a2}} which consists of $>100$~million galaxies to $i<23.5$~mag \citep{SevillaNoarbe2021,Gatti_2021}.  
DES Y3 ellipticities are measured via \textsc{Metacalibration}, in which the shape measurement biases are inferred by artificially shearing the observed images \citep{Huff2017, Sheldon2017, Gatti_2021}.
The tomographic redshift distributions are calibrated following \cite{Myles2021} and the shear calibration is from \cite{maccrann_2022}.
We also consider the DES~Y3 ``blue shear" catalog of \cite{McCullough2024}, which omits red source galaxies (Appendix~\ref{app:xray_sensitivity}).

The Kilo Degree Survey (KiDS; \citealp{dejong2013}) was carried out at Paranal from 2011 to 2019 using OmegaCAM on the VLT Survey Telescope. 
We consider the fourth data release \citep[DR4\footnote{\url{https://kids.strw.leidenuniv.nl/DR4/index.php}},][]{kuijken2019} which covers approximately $1000\,{\rm deg}^2$. 
Shapes are measured with the \textit{lens}fit algorithm \citep{Miller2013, FenechConti2017} and validated in \citet{Giblin_2021}.
Photometric redshifts are calculated with self-organizing maps \citep{Wright2020,Hildebrandt2020}.

The Hyper Suprime-Cam (HSC) Subaru Strategic Program is conducting a 300-night $grizy$ imaging survey with the $8.2$~meter Subaru telescope \citep{Aihara2018}.
We use the three year release,\footnote{\url{https://hsc-release.mtk.nao.ac.jp/doc/}} which covers $416$~deg$^2$ to $i < 24.5$~mag \citep{Li2022}.
Shapes are measured with the \texttt{GalSim} \citep{Rowe2015} re-Gaussianization PSF correction method \citep{Hirata2003}.
The redshift distributions are derived using the \texttt{DNNz} neural network estimator \citep{Rau2023}.
We correct the mean redshifts of the HSC tomographic bins following \cite{Li2022} and \cite{Heydenreich2025}.

\subsection{Lensing Corrections}
\label{app:lensing_corrections}

For each lensing survey, we correct the raw excess surface density estimator (Equation~\ref{eqn:raw_esd}) for known systematics following \cite{Amon2023} and \cite{Heydenreich2025}, which we briefly outline below.
In Appendix~\ref{appendix:sensitivity}, we report the sensitivity of our halo mass estimates to these corrections.

\subsubsection{Shear bias corrections}

The measured galaxy shapes must be calibrated to yield accurate tangential shear estimates.
Shear calibration is primarily multiplicative and varies between imaging surveys.
Shear calibration is applied for all our GGL measurements.

For DES~Y3, shape measurement biases are inferred by artificially shearing the observed images \citep[\textsc{Metacalibration},][]{Huff2017, Sheldon2017, Gatti_2021}.
A shear response matrix $\bm{\mathcal{R}}$ is calculated from the artificially sheared images for each galaxy:
\begin{equation}
    \bm{\mathcal{R}}_{i,j} = \frac{ \epsilon_i^{ s_{j+} } - \epsilon_i^{ s_{j-} } }{ \Delta \gamma_j },
\end{equation}
where $\epsilon_i^{ s_{j+/-} }$  is the measured ellipticity of the $i$th component on the image positively (negatively) sheared by $ \Delta \gamma_j$ in the $j$th component.
The full response matrix is well approximated by $\mathcal{R} = ( \bm{\mathcal{R}}_{11} + \bm{\mathcal{R}}_{22} )/2$ \citep[see Appendix A of][]{Gatti_2021}.
We therefore approximate the average responsivity as
\begin{equation}
    \mathcal{R} = \frac{\sum_{i=1}^N  w_i (\bm{\mathcal{R}}_{i,11}+\bm{\mathcal{R}}_{i,22})/2 }{ \sum_{i=1}^N  w_i },
\end{equation}
where $w_i$ is the inverse-variance weight of the $i$th galaxy and $N$ is the number of galaxies.
Additional biases (e.g., shear-dependent detection and blending) are calibrated by image simulations and corrected with the multiplicative biases $m$ \citep{maccrann_2022}.
For DES, the shear calibrated excess surface density estimator is
\begin{equation}
    \Delta \Sigma (R) = \frac{1}{(1+m)\mathcal{R}} \Delta \Sigma_\mathrm{raw} (R).
\end{equation}
For the DES~Y3 blue shear catalog, \cite{McCullough2024} adopted the \textsc{Metacalibration}  per-galaxy response measurements and recalibrated the multiplicative bias following \cite{maccrann_2022}. 

KiDS~1000 employs a self-calibrating variant of the \textit{lens}fit algorithm \citep{Miller2013,FenechConti2017}.
The multiplicative bias $m$ was calculated per tomographic bin \citep{Hildebrandt2020,Giblin_2021}.
For KiDS, the shear calibrated excess surface density estimator is
\begin{equation}
    \Delta \Sigma (R) = \frac{1}{(1+m)} \Delta \Sigma_\mathrm{raw} (R).
\end{equation}

For HSC~Y3 the shear responsivity is approximated as
\begin{equation}
    \mathcal{R} = 1 - \frac{ \sum_{i=1}^{N} w_i e_\mathrm{i,rms}^2 }{ \sum_{i=1}^{N} w_i },
\end{equation}
where $e_{i,\mathrm{rms}}$ is the RMS of the intrinsic ellipticity for the $i$th galaxy and $w_i$ is the per-galaxy inverse-variance weight \citep{Hirata2003}.
Residual multiplicative biases are estimated as a function of signal-to-noise and resolution, using Hubble Space Telescope observations overlapping the HSC footprint \citep{Li2022}.
We adopt the weighted average multiplicative bias across the shear catalog. 
HSC~Y3 also corrects for multiplicative and additive biases induced by the selection criteria for sources using image simulations \citep{More2023,Li2023}.
The HSC shear calibrated excess surface density estimator is then
\begin{equation}
    \Delta \Sigma (R) = \frac{1}{1+m_\mathrm{sel}} \left (\frac{1}{2 \mathcal{R}(1+m)} \Delta \Sigma_\mathrm{raw} (R)  - a_\mathrm{sel} \Delta \Sigma^\mathrm{psf} \right),
\end{equation}
where $m_\mathrm{sel}$ is the multiplicative selection bias,  $a_\mathrm{sel}$ is the additive selection bias, and $\Delta \Sigma^\mathrm{psf}$ is the weighted average PSF ellipticity $e^\mathrm{psf}$ across the source positions: 
\begin{equation}
     \Delta \Sigma^\mathrm{psf} =  \frac{ \sum_{i=1}^{N} w_i e_i^\mathrm{psf} }{ \sum_{i=1}^{N} w_i }.
\end{equation}

\subsubsection{Randoms subtraction}

Large-scale structure also imprints tangential shear on the source galaxies, in addition to the shear induced by the lenses. 
To isolate the GGL signal and account for residual additive bias, we compute the excess surface density around randomly distributed positions on the sky $\Delta \Sigma_\mathrm{random}$ and subtract it from the raw excess surface density estimator for the DESI lenses.
For DESI, we employ the DR1 large scale structure random catalogs.
The randoms sample the area on the sky where DESI~DR1 data could have been observed and reproduce the redshift distribution of the observations by randomly drawing redshifts directly from the observed data \citep{Ross2025}.
We forgo the randoms correction for the eROSITA cluster measurements, because no randoms catalog is readily available. 
Because the overlap of the eROSITA footprint with the imaging surveys is considerably larger than for DESI (Figure~\ref{fig:ggl_comparison}), the contamination from large-scale structure is expected to be minor; even for DESI lenses, $\Delta \Sigma_\mathrm{random}$ is consistent with zero.

\subsubsection{Boost factor correction}

Because the photometric redshifts of sources are uncertain, the source redshift distribution can overlap with the lenses.
Galaxies that are physically close to lenses but are erroneously identified as sources will bias the lensing signal low, particularly at small transverse separations.

To account for this bias, we multiply the raw excess surface density estimator by the boost factor $B(R)$, which estimates the amount of excess sources behind a lens compared with the random sky positions \citep{Sheldon2004}:
\begin{equation}
    B(R) = \frac{\sum_\mathrm{ls} w_\mathrm{l} w_\mathrm{s} }{\sum_\mathrm{rs} w_\mathrm{r} w_\mathrm{s}},
\end{equation}
where the numerator sums the product of the lens--source pair weights ($w_\mathrm{l} w_\mathrm{s}$) and the denominator sums the product of the random--source pair weights ($w_\mathrm{r} w_\mathrm{s}$);
we again draw the randoms from the DESI~DR1 randoms catalogs \citep{Ross2025}.

The boost correction is only applied for the DESI lenses, which are at higher redshift than the eROSITA lenses. 
For the eROSITA clusters, although the lenses are typically better separated from the sources in redshift, the cluster members are predominantly red and quenched, which increases the likelihood that cluster members are mistakenly identified as background source galaxies due to photometric redshift uncertainties.
We correct for cluster lens--source contamination in Appendix~\ref{app:xray_sensitivity}.

\subsubsection{Lens magnification bias}

Lensing by large scale structure induces a (de)magnification effect on galaxy fluxes (in addition to shear).
Areas of the sky with high magnification by foreground over densities will therefore have higher number densities of detected galaxies.
This connection between detection probability and large scale structure biases excess surface density measurements: lenses are more likely to be detected if they are magnified by foreground structure, and that same structure shears the background source galaxies.
We measure the magnification bias $\Delta \Sigma_\mathrm{mag}$ with the \texttt{dsigma} implementation of \cite{Unruh2020};
we model the matter power spectrum with \texttt{CAMB}\footnote{\url{https://camb.readthedocs.io}} and set $A_s=2.83\times10^{-9}$.
The impact of the magnification bias depends on the faint-end slope of the galaxy luminosity function $\alpha$:
\begin{equation}
    \alpha = \frac{ \mathrm{d} \ln f }{ \mathrm{d} \mu }  \Big |_{\mu=1},
\end{equation}
where $f$ is the fraction of galaxies passing a magnitude selection and $\mu$ is the magnification amplitude.
\cite{Heydenreich2025} report $\alpha$ for DESI BGS and LRG in three redshift bins. 
We adopt $\alpha_\mathrm{BGS}=2.19$ and $\alpha_\mathrm{LRG}= 2.52$ by interpolating the reported $\alpha$ at the mean redshift of the BGS and LRG samples.
Following \cite{Amon2023}, we adopt $\alpha_\mathrm{LOWZ}=2.19$ and $\alpha_\mathrm{CMASS}= 2.52$ from \cite{vonWietersheimKramsta2021} for the SDSS samples.

\subsubsection{Measurement Covariances and Scale Cuts}
\label{app:scale_cuts}

We estimate the covariance matrix of the GGL measurements with a two-dimensional leave-one-out Jackknife process.
For each lens sample, we divide the on sky footprint into $100$ patches via K-means clustering using \texttt{dsigma} \citep{Lange2022}.
We successively measure the excess surface density with one patch left out at a time. 
The mean GGL signal and the covariance matrix are then calculated from the collection of leave-one-out measurements.

The covariance matrix is only valid on angular scales smaller than the Jackknife patches \citep[e.g., Appendix A of][]{Johnston2019}.
For the DESI lenses, we consider three imaging surveys (DES, KiDS, and HSC).
The Jackknife patches are smallest for the HSC shear catalog: $36.5'$ across, corresponding to $9$ and $20$ comoving~{\Mpch} at the mean redshifts of the BGS and LRG lenses, respectively.
For the eROSITA lenses, we rely on the DES shear catalog because of its considerably larger overlap with the eROSITA footprint (Figure~\ref{fig:photoksz}).
The DES Jackknife patches are $110'$ across, corresponding to $14$ comoving~{\Mpch} at $z=0.15$ (the approximate redshift of our eROSITA lens sample).

\section{Selecting a like-for-like simulated sample}
\label{appendix:sensitivity}

In this paper, we perform like-with-like comparisons between kSZ and X-ray measurements and the FLAMINGO simulations. 
For each sample, we identify the selection of FLAMINGO halos that best-fits the measured GGL profile.
We discuss the sensitivity of our results to the method of selecting simulated galaxies and corrections to the GGL for the kSZ and X-ray samples in Appendices~\ref{app:ksz_sensitivity} and \ref{app:xray_sensitivity}, respectively.

\subsection{SDSS and DESI kSZ samples}
\label{app:ksz_sensitivity}

For the kSZ samples, our fiducial method of selecting simulated galaxies (Section~\ref{sec:matching_ksz}) is to select all galaxies above a minimum stellar mass threshold \citep{McCarthy2025};
we find the stellar mass bound that yields the best-fit between the FLAMINGO $\Delta \Sigma$ profile and the observed GGL signal.
This selection includes both centrals and satellites. 
To account for the redshift distribution of the observed sample, the simulated $\Delta \Sigma$ profile is a weighted average across the FLAMINGO redshift shells.
We repeat the fitting process for each FLAMINGO simulation.

To understand the robustness of the calibration, 
we explore multiple variations.
We select galaxies in terms of halo mass, instead of stellar mass; 
the halo mass selection is limited to centrals, while the stellar mass cut includes both centrals and satellites. 
We also select galaxies from a log-normal distribution of either stellar mass or halo mass. 
For the log-normal selection, we fit for the central mass when calibrating to the observed GGL profile; we fix the standard deviation to $0.2$~dex.
For each variation of the calibration procedure, we report the mean halo mass of the selected FLAMINGO galaxies, as well as the goodness of fit between the simulation and the observed GGL and kSZ profiles; see Tables~\ref{tab:desi_summary} and \ref{tab:sdss_summary} for the DESI and SDSS samples, respectively.
For brevity, we only report the halo masses derived from the fiducial and strongest feedback simulations.
The goodness of fits between the simulated kSZ signals and data are weakly sensitive to the above variations.

We also investigated how sensitive the inferred halo masses are to variations in the GGL measurement. 
As outlined in Appendix~\ref{app:lensing_corrections}, we apply several corrections to the excess surface density measurement. 
The largest correction is the shear bias correction; for modern lensing surveys, shape measurements are calibrated to better than $1\%$ \citep[e.g.,][]{Amon2023,Li2023}, which corresponds to $<0.005$~dex uncertainty in the mean halo mass.
All other corrections (randoms subtraction, boost factor, lens magnification bias) shift the GGL signal by $<10\%$ collectively (i.e., $<0.05$~dex in halo mass).

\subsection{eROSITA cluster samples}
\label{app:xray_sensitivity}

For the eROSITA lens samples, our fiducial method of selecting simulated halos (Section~\ref{sec:matching_xray}) is to select all centrals above a given mass.
For each bin of eROSITA clusters, we identify the minimum halo mass cutoff such that the simulated $\Delta \Sigma$ profile matches the observed GGL signal. 
The simulated $\Delta \Sigma$ profiles are weighted averages over the FLAMINGO lightcone redshift shells, where the weights are set by the redshift distributions of the eROSITA cluster samples. 

Analogous to the kSZ samples (Appendix~\ref{app:ksz_sensitivity}), we explore variations of the GGL fitting process.
We repeat the fit for each FLAMINGO simulation and find that the derived halo masses are insensitive to the simulation's feedback strength ($<0.02$~dex).
Instead of selecting halos above a minimum mass, we draw halos from a log-normal distribution of halo mass. 
We then fit for the central mass that best reproduces the observed GGL profile; we again fix the standard deviation to $0.2$~dex.
For the higher redshift bin ($z=0.1-0.2$), we consider both the primary eRASS1 sample and the higher purity ``cosmology" sample, which requires greater detection probability \citep{Bulbul2024}.
In Table~\ref{tab:eROSITA_samples}, we report the derived halo masses for each variation and find that the halo masses are insensitive to the changes ($<0.05$~dex).
For brevity, we only report the halo masses from the fiducial feedback FLAMINGO simulation.

GGL measurements of galaxy clusters are challenged by the fact that cluster members can erroneously be included in the background source galaxy sample due to photometric redshift uncertainties, which dilutes the shear signal \citep{Hoekstra2012,Gruen2014,Dietrich2019,Varga2019}.
Cluster member contamination can be modeled as a function of redshift, richness, and separation from the cluster center \citep[e.g.,][]{Grandis2024}, 
however, these corrections require careful calibration against simulations.
Instead, we adopt several strategies to minimize contamination at the data level:
we restrict our analysis to larger scales \citep[$>2$ comoving {\Mpch},][]{Kleinebreil2025}, and we repeat the GGL measurement using the  blue shear catalog of \cite{McCullough2024}.
Because cluster members are predominantly red and quenched, blue galaxies are less likely to be cluster galaxies erroneously classified as background sources \citep{Chiu2022}.
In Table~\ref{tab:eROSITA_samples}, we report the derived halo masses using the fiducial DES~Y3 shear catalog and the blue shear catalog. The inferred halo masses are insensitive to the choice of shear catalog ($<0.05$~dex change in mean halo mass).

\begin{figure*}[t!]
\includegraphics[width=\textwidth]{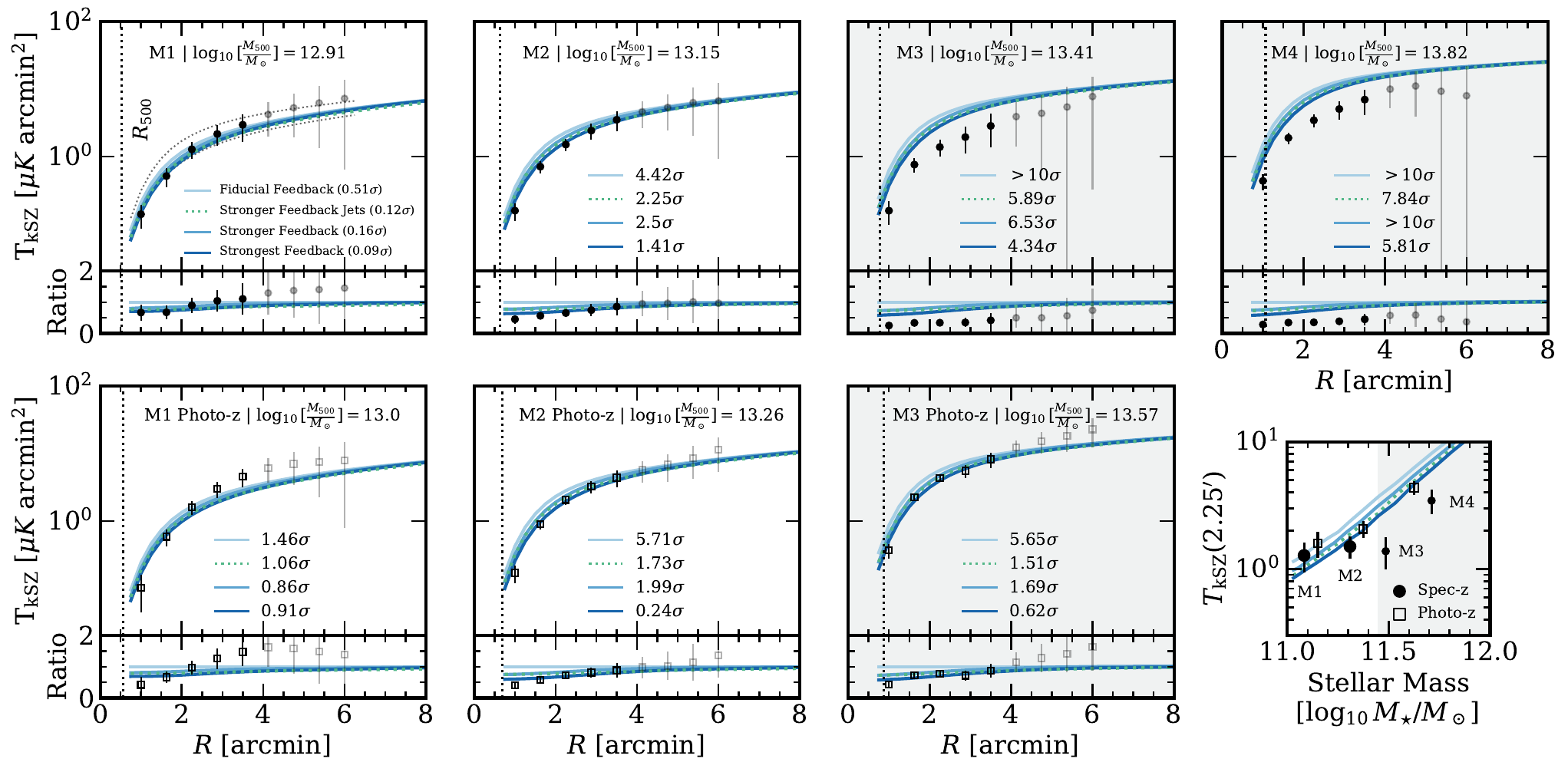}
\caption{
Comparison of the photometric (bottom) and spectroscopic (top) kSZ measurements for LRG stellar mass bins. 
The stacked kSZ profiles for the selection of simulated galaxies that best-fits the GGL are shown for the four FLAMINGO simulations we consider.
We report the number of standard deviations that the simulation predictions deviate from the observations.
For the spectroscopic kSZ measurements, we calculate goodness of fit using the full measurement covariance matrices. For the photometric measurements, only diagonal uncertainties are readily available, so we restrict the goodness of fit calculation to $\theta<4'$ (the outermost radial bins are highly correlated).  
The lower panels show the ratio of the data and the simulation predictions to the fiducial simulation. 
The vertical dotted line demarcates $R_{500}$.
The kSZ effect measurement is highly correlated at larger angular separations, so these bins are shaded light gray to reflect their lower statistical weight.
The lower right panel presents the amplitude of the stacked kSZ effect at $\theta=2.25'$ for the spectroscopic \citep{ReidkSZ} and photometric \citep{Hadzhiyska2024} LRG mass bins as a function of mean stellar mass \citep{Zhou2023LRG}, alongside the simulation predictions.
At $M_{\star}>2.5 \times 10^{11}~M_\odot$ (shaded gray), the spectroscopic and photometric results require different feedback strengths.
}
\label{fig:photoksz}
\end{figure*}

\section{Analysis of photometric kSZ}
\label{appendix:photo-ksz}

Measurement of the kSZ effect relies upon reconstructed peculiar velocities.
These velocities can be derived from spectroscopic or photometric redshifts; 
while photometric redshifts are more uncertain, the measured kSZ effect profile can reach comparable signal-to-noise to a spectroscopic measurement if the sample is large enough \citep{RiedGuachalla2024,Hadzhiyska2024}.
In our primary analysis, we consider the spectroscopic based kSZ measurements of \cite{Schaan2021} and \cite{ReidkSZ}.
Complementing these measurements, \cite{Hadzhiyska2024photoz} used photometric redshifts from the DESI Legacy Imaging Survey \citep[DR9 and DR10,][]{Zhou2023LRG} to measure the kSZ effect for LRGs, also using the ACT~DR6 map.

\cite{Hadzhiyska2024photoz} report the photometric based kSZ effect profile for four stellar mass bins: $\log_{10} M_*/M_\odot=(11,11.25),(11.25,11.5),(11.5,12)$, and $(12,13.5)$.
We omit the highest mass bin from our analysis, because it contains only $2,000$ galaxies and is potentially contaminated by the tSZ effect.
We measure the GGL profile for each bin and derive a mean halo mass from the selection of simulated FLAMINGO galaxies that best-fits the GGL; we use our fiducial selection function of a minimum stellar mass cut, which includes both centrals and satellites.
In Figure~\ref{fig:ksz_result}, we present the amplitude of the kSZ effect profile at $2.25'$ as a function of GGL derived halo mass for the spectroscopic and photometric based kSZ measurements.
Below $M_{500} = 2 \times 10^{13}~M_\odot$, the spectroscopic and photometric kSZ measurements are both well described by the strongest feedback FLAMINGO simulation.
At higher masses, the spectroscopic measurements require relatively stronger feedback than the photometric measurements.
This disagreement is independent of our GGL halo masses. 
In Figure~\ref{fig:photoksz}, we present the amplitude of the kSZ effect profile at $2.25'$ as a function of mean stellar mass, using the photometric based stellar masses of \cite{Zhou2023LRG}.
In terms of stellar mass, the spectroscopic measurements still require stronger feedback than the photometric measurements for the highest mass bins.

The kSZ amplitude versus halo mass relation neglects the full shape of the kSZ effect profile.
We therefore perform our like-with-like analysis on the photometric based kSZ profiles.
For each stellar mass bin, we calculate the stacked kSZ effect profile from the selection of simulated galaxies that best-fits the GGL measurements; 
following our primary analysis, we marginalize over the redshift distribution of each stellar mass bin.
The simulation predictions are presented against the data in Figure~\ref{fig:photoksz}. 
We also present the like-with-like comparison for all four spectroscopic based LRG kSZ bins.
In Figure~\ref{fig:photoksz}, we report the number of standard deviations between the data and the simulation predictions.
For the spectroscopic measurements, we use the full covariance matrix, but for the photometric measurements, only diagonal uncertainties are readily available; for the photometric measurements, the goodness of fit is therefore calculated with the diagonal errors for $\theta<4'$ (at larger separations the measurements are highly correlated). 
Our simulation comparison finds that the photometric kSZ measurements are well described by the strongest feedback FLAMINGO simulation ({\eightsigma}) for all mass bins, as are the two lowest mass bins of the spectroscopic measurements (M1 and M2).
However, the two highest mass spectroscopic bins (M3 and M4) appear to require even stronger feedback than the photometric measurements.

Given the differences between the spectroscopic and photometric based kSZ measurements, we omit the highest mass LRG bins from our primary analysis. 
Both the spectroscopic and photometric measurements require stronger feedback than the fiducial FLAMINGO simulation, but they differ on whether even stronger feedback than {\eightsigma} is required.
Future work is necessary to understand this result, including an investigation of observational uncertainties and potential systematics in how we match with simulations.

\begin{deluxetable*}{ccccccc}[h]
\tablecaption{DESI GGL Calibration Summary.\label{tab:desi_summary}}
\tablehead{
 \colhead{Name} & \colhead{Simulation} & \colhead{Satellites} & \colhead{Selection} & \colhead{Halo Mass} & \colhead{GGL $\chi^2_{\nu}$} & \colhead{kSZ $\chi^2_{\nu}$}
}
\startdata
BGS & Fiducial Feedback &  $\checkmark$ & One sided & $13.31 \pm 0.009$ & $0.98~(0.77\sigma)$ & $2.14~(2.18\sigma)$ \\
 &  &  $\checkmark$ & Log--normal & $13.3 \pm 0.008$ & $0.76~(0.49\sigma)$ & $2.01~(2.04\sigma)$ \\
 &  &    & One sided & $13.09 \pm 0.01$ & $8.46~(6.35\sigma)$ & $2.09~(2.14\sigma)$ \\
 &  &    & Log--normal & $12.99 \pm 0.02$ & $12.2~(8.01\sigma)$ & $1.66~(1.63\sigma)$ \\
\cline{2-7}
 & Strongest Feedback &  $\checkmark$ & One sided & $13.31 \pm 0.01$ & $1.01~(0.8\sigma)$ & $1.08~(0.89\sigma)$ \\
 &  &  $\checkmark$ & Log--normal & $13.3 \pm 0.009$ & $1.0~(0.8\sigma)$ & $1.03~(0.82\sigma)$ \\
 &  &    & One sided & $13.09 \pm 0.01$ & $8.05~(6.15\sigma)$ & $0.99~(0.78\sigma)$ \\
 &  &    & Log--normal & $13.0 \pm 0.02$ & $11.96~(7.93\sigma)$ & $0.9~(0.65\sigma)$ \\
\hline
LRG M1 & Fiducial Feedback &  $\checkmark$ & One sided & $12.91 \pm 0.02$ & $1.3~(1.19\sigma)$ & $0.79~(0.51\sigma)$ \\
 &  &  $\checkmark$ & Log--normal & $12.91 \pm 0.01$ & $1.27~(1.16\sigma)$ & $0.82~(0.55\sigma)$ \\
 &  &    & One sided & $12.69 \pm 0.02$ & $3.82~(3.95\sigma)$ & $0.66~(0.35\sigma)$ \\
 &  &    & Log--normal & $12.62 \pm 0.02$ & $4.4~(4.44\sigma)$ & $0.52~(0.2\sigma)$ \\
\cline{2-7}
 & Strongest Feedback &  $\checkmark$ & One sided & $12.92 \pm 0.01$ & $1.15~(0.99\sigma)$ & $0.39~(0.09\sigma)$ \\
 &  &  $\checkmark$ & Log--normal & $12.91 \pm 0.02$ & $1.25~(1.13\sigma)$ & $0.39~(0.09\sigma)$ \\
 &  &    & One sided & $12.69 \pm 0.02$ & $3.65~(3.8\sigma)$ & $0.46~(0.15\sigma)$ \\
 &  &    & Log--normal & $12.61 \pm 0.02$ & $4.14~(4.22\sigma)$ & $0.52~(0.2\sigma)$ \\
\hline
LRG M2 & Fiducial Feedback &  $\checkmark$ & One sided & $13.15 \pm 0.009$ & $1.78~(1.84\sigma)$ & $4.68~(4.42\sigma)$ \\
 &  &  $\checkmark$ & Log--normal & $13.14 \pm 0.009$ & $1.97~(2.07\sigma)$ & $4.72~(4.45\sigma)$ \\
 &  &    & One sided & $13.01 \pm 0.009$ & $8.82~(7.34\sigma)$ & $3.87~(3.81\sigma)$ \\
 &  &    & Log--normal & $12.92 \pm 0.02$ & $9.99~(7.96\sigma)$ & $3.45~(3.45\sigma)$ \\
\cline{2-7}
 & Strongest Feedback &  $\checkmark$ & One sided & $13.15 \pm 0.01$ & $1.47~(1.43\sigma)$ & $1.48~(1.41\sigma)$ \\
 &  &  $\checkmark$ & Log--normal & $13.14 \pm 0.01$ & $1.56~(1.55\sigma)$ & $1.53~(1.47\sigma)$ \\
 &  &    & One sided & $13.01 \pm 0.009$ & $8.36~(7.08\sigma)$ & $1.33~(1.22\sigma)$ \\
 &  &    & Log--normal & $12.92 \pm 0.02$ & $9.53~(7.72\sigma)$ & $1.22~(1.08\sigma)$ \\
\hline
LRG M3 & Fiducial Feedback &  $\checkmark$ & One sided & $13.41 \pm 0.009$ & $2.23~(2.37\sigma)$ & $15.21~(>10\sigma)$ \\
 &  &  $\checkmark$ & Log--normal & $13.41 \pm 0.008$ & $2.41~(2.58\sigma)$ & $15.63~(>10\sigma)$ \\
 &  &    & One sided & $13.3 \pm 0.01$ & $7.33~(6.48\sigma)$ & $14.87~(>10\sigma)$ \\
 &  &    & Log--normal & $13.24 \pm 0.01$ & $8.09~(6.93\sigma)$ & $14.83~(>10\sigma)$ \\
\cline{2-7}
 & Strongest Feedback &  $\checkmark$ & One sided & $13.41 \pm 0.01$ & $1.83~(1.9\sigma)$ & $4.56~(4.34\sigma)$ \\
 &  &  $\checkmark$ & Log--normal & $13.41 \pm 0.008$ & $1.81~(1.88\sigma)$ & $5.08~(4.72\sigma)$ \\
 &  &    & One sided & $13.3 \pm 0.01$ & $6.46~(5.93\sigma)$ & $4.02~(3.92\sigma)$ \\
 &  &    & Log--normal & $13.24 \pm 0.01$ & $7.13~(6.35\sigma)$ & $3.73~(3.68\sigma)$ \\
\hline
LRG M4 & Fiducial Feedback &  $\checkmark$ & One sided & $13.82 \pm 0.008$ & $4.91~(4.84\sigma)$ & $22.01~(>10\sigma)$ \\
 &  &  $\checkmark$ & Log--normal & $13.82 \pm 0.01$ & $3.94~(4.06\sigma)$ & $21.29~(>10\sigma)$ \\
 &  &    & One sided & $13.78 \pm 0.01$ & $7.48~(6.57\sigma)$ & $22.06~(>10\sigma)$ \\
 &  &    & Log--normal & $13.73 \pm 0.01$ & $7.27~(6.44\sigma)$ & $22.55~(>10\sigma)$ \\
\cline{2-7}
 & Strongest Feedback &  $\checkmark$ & One sided & $13.81 \pm 0.009$ & $3.73~(3.88\sigma)$ & $6.77~(5.81\sigma)$ \\
 &  &  $\checkmark$ & Log--normal & $13.81 \pm 0.01$ & $3.05~(3.25\sigma)$ & $7.37~(6.16\sigma)$ \\
 &  &    & One sided & $13.77 \pm 0.01$ & $5.97~(5.6\sigma)$ & $6.47~(5.62\sigma)$ \\
 &  &    & Log--normal & $13.73 \pm 0.01$ & $5.86~(5.52\sigma)$ & $6.84~(5.85\sigma)$ \\
\enddata
\tablecomments{For each DESI kSZ sample, we report the average halo mass $M_{500}$ inferred from the GGL FLAMINGO calibration, as well as the goodness of fit between the simulation and the observed GGL and kSZ profiles.
We consider multiple variations of the GGL calibration, including i) selecting FLAMINGO galaxies in terms of halo mass (centrals only) or stellar mass (centrals and satellites), ii) applying a one sided selection (all galaxies above a given mass) or selecting galaxies according to a log-normal distribution at a given central mass, and iii) fitting to the fiducial or strongest feedback ({\eightsigma}) FLAMINGO simulations. 
}
\end{deluxetable*}

\begin{deluxetable*}{cccccccc}[h]
\tablecaption{LOWZ and CMASS GGL Calibration Summary.\label{tab:sdss_summary}}
\tablehead{
 \colhead{Name} & \colhead{Simulation} & 
 \colhead{$n(z)$ marginalized} & \colhead{Satellites} & \colhead{Selection} & \colhead{Halo Mass} & \colhead{GGL $\chi^2_{\nu}$} & \colhead{kSZ $\chi^2_{\nu}$}
}
\startdata
LOWZ & Fiducial Feedback &    & $\checkmark$ & One sided & $13.5 \pm 0.02$ & $0.28~(0.02\sigma)$ & $8.05~(6.54\sigma)$ \\
 &  &  $\checkmark$ & $\checkmark$ & One sided & $13.51 \pm 0.01$ & $0.4~(0.07\sigma)$ & $5.54~(5.03\sigma)$ \\
 &  &    & $\checkmark$ & Log--normal & $13.51 \pm 0.02$ & $0.34~(0.04\sigma)$ & $9.03~(7.05\sigma)$ \\
 &  &  $\checkmark$ & $\checkmark$ & Log--normal & $13.51 \pm 0.02$ & $0.38~(0.06\sigma)$ & $6.16~(5.43\sigma)$ \\
 &  &    &   & One sided & $13.38 \pm 0.02$ & $1.16~(1.01\sigma)$ & $11.86~(>10\sigma)$ \\
 &  &  $\checkmark$ &   & One sided & $13.37 \pm 0.02$ & $1.47~(1.46\sigma)$ & $8.31~(6.67\sigma)$ \\
 &  &    &   & Log--normal & $13.31 \pm 0.02$ & $1.5~(1.5\sigma)$ & $12.12~(>10\sigma)$ \\
 &  &  $\checkmark$ &   & Log--normal & $13.3 \pm 0.02$ & $1.8~(1.92\sigma)$ & $8.55~(6.8\sigma)$ \\
\cline{2-8}
 & Strongest Feedback &    & $\checkmark$ & One sided & $13.49 \pm 0.02$ & $0.42~(0.08\sigma)$ & $1.39~(1.29\sigma)$ \\
 &  &  $\checkmark$ & $\checkmark$ & One sided & $13.5 \pm 0.01$ & $0.49~(0.12\sigma)$ & $1.11~(0.92\sigma)$ \\
 &  &    & $\checkmark$ & Log--normal & $13.5 \pm 0.03$ & $0.37~(0.05\sigma)$ & $1.33~(1.22\sigma)$ \\
 &  &  $\checkmark$ & $\checkmark$ & Log--normal & $13.5 \pm 0.02$ & $0.44~(0.09\sigma)$ & $1.05~(0.85\sigma)$ \\
 &  &    &   & One sided & $13.36 \pm 0.02$ & $0.85~(0.56\sigma)$ & $2.06~(2.1\sigma)$ \\
 &  &  $\checkmark$ &   & One sided & $13.35 \pm 0.02$ & $1.02~(0.8\sigma)$ & $1.44~(1.36\sigma)$ \\
 &  &    &   & Log--normal & $13.31 \pm 0.02$ & $1.4~(1.36\sigma)$ & $1.89~(1.91\sigma)$ \\
 &  &  $\checkmark$ &   & Log--normal & $13.29 \pm 0.02$ & $1.46~(1.45\sigma)$ & $1.45~(1.37\sigma)$ \\
\hline
CMASS & Fiducial Feedback &    & $\checkmark$ & One sided & $13.29 \pm 0.02$ & $2.25~(2.39\sigma)$ & $5.29~(4.86\sigma)$ \\
 &  &  $\checkmark$ & $\checkmark$ & One sided & $13.29 \pm 0.02$ & $2.29~(2.44\sigma)$ & $4.89~(4.58\sigma)$ \\
 &  &    & $\checkmark$ & Log--normal & $13.28 \pm 0.03$ & $2.12~(2.25\sigma)$ & $6.34~(5.54\sigma)$ \\
 &  &  $\checkmark$ & $\checkmark$ & Log--normal & $13.27 \pm 0.03$ & $2.09~(2.21\sigma)$ & $6.01~(5.33\sigma)$ \\
 &  &    &   & One sided & $13.16 \pm 0.03$ & $1.14~(0.97\sigma)$ & $7.65~(6.31\sigma)$ \\
 &  &  $\checkmark$ &   & One sided & $13.16 \pm 0.03$ & $1.17~(1.02\sigma)$ & $7.57~(6.27\sigma)$ \\
 &  &    &   & Log--normal & $13.1 \pm 0.03$ & $1.42~(1.36\sigma)$ & $6.92~(5.9\sigma)$ \\
 &  &  $\checkmark$ &   & Log--normal & $13.1 \pm 0.03$ & $1.55~(1.53\sigma)$ & $6.46~(5.62\sigma)$ \\
\cline{2-8}
 & Strongest Feedback &    & $\checkmark$ & One sided & $13.29 \pm 0.03$ & $2.18~(2.32\sigma)$ & $1.01~(0.8\sigma)$ \\
 &  &  $\checkmark$ & $\checkmark$ & One sided & $13.29 \pm 0.03$ & $2.3~(2.45\sigma)$ & $1.03~(0.83\sigma)$ \\
 &  &    & $\checkmark$ & Log--normal & $13.26 \pm 0.03$ & $2.29~(2.45\sigma)$ & $1.09~(0.9\sigma)$ \\
 &  &  $\checkmark$ & $\checkmark$ & Log--normal & $13.26 \pm 0.03$ & $2.29~(2.44\sigma)$ & $1.07~(0.88\sigma)$ \\
 &  &    &   & One sided & $13.13 \pm 0.03$ & $1.29~(1.19\sigma)$ & $1.07~(0.88\sigma)$ \\
 &  &  $\checkmark$ &   & One sided & $13.14 \pm 0.03$ & $1.37~(1.29\sigma)$ & $1.11~(0.92\sigma)$ \\
 &  &    &   & Log--normal & $13.1 \pm 0.03$ & $1.65~(1.67\sigma)$ & $1.01~(0.8\sigma)$ \\
 &  &  $\checkmark$ &   & Log--normal & $13.1 \pm 0.03$ & $1.53~(1.51\sigma)$ & $1.04~(0.84\sigma)$ \\
\enddata
\tablecomments{For each SDSS galaxy sample, we report the average halo mass $M_{500}$ inferred from the GGL FLAMINGO calibration, as well as the goodness of fit to observations.
We consider the same variations to the calibration as Table~\ref{tab:desi_summary} but include calibrations without marginalizing over the observed sample's redshift distribution $n(z)$ to facilitate comparison with \cite{McCarthy2025}.
}
\end{deluxetable*}

\begin{deluxetable*}{ccccccc}
\tablecaption{eROSITA GGL Calibration.\label{tab:eROSITA_samples}}
\tablehead{
 \colhead{Redshift Selection} & \colhead{Halo Mass Selection} & \colhead{eRASS1 Catalog} & \colhead{Shape Catalog} & \colhead{Selection} & \colhead{eRASS1 Mass} & \colhead{GGL Mass} \\
 \colhead{} & \colhead{$\log_{10}[  M_{500}/M_\odot ] $} & \colhead{} & \colhead{} & \colhead{} & \colhead{$\log_{10}[\langle M_{500}/M_\odot\rangle] $} & \colhead{$\log_{10}[\langle M_{500}/M_\odot\rangle] $}
}
\startdata
$0.05-0.1$ & $13.3-14.0$ & Primary  & Fiducial & One sided & $13.733\pm0.009$ & $13.71\pm0.1$\\
 &  &   &  & Log--normal &  & $13.66\pm0.09$\\
 &  &   & Blue shear & One sided &  & $13.69\pm0.12$\\
 &  &   &  & Log--normal &  & $13.65\pm0.1$\\
\cline{2-7}
 & $14.0-14.5$ & Primary  & Fiducial & One sided & $14.255\pm0.004$ & $14.19\pm0.05$\\
 &  &   &  & Log--normal &  & $14.14\pm0.04$\\
 &  &   & Blue shear & One sided &  & $14.14\pm0.07$\\
 &  &   &  & Log--normal &  & $14.12\pm0.05$\\
\hline
$0.1-0.2$ & $13.5-14.0$ & Primary  & Fiducial & One sided & $13.804\pm0.009$ & $13.84\pm0.07$\\
 &  &   &  & Log--normal &  & $13.78\pm0.06$\\
 &  &   & Blue shear & One sided &  & $13.83\pm0.09$\\
 &  &   &  & Log--normal &  & $13.77\pm0.08$\\
\cline{3-7}
 &  & Cosmology  & Fiducial & One sided & $13.844\pm0.011$ & $13.9\pm0.04$\\
 &  &   &  & Log--normal &  & $13.84\pm0.08$\\
 &  &   & Blue shear & One sided &  & $13.94\pm0.06$\\
 &  &   &  & Log--normal &  & $13.91\pm0.1$\\
\cline{2-7}
 & $14.0-14.5$ & Primary  & Fiducial & One sided & $14.237\pm0.004$ & $14.19\pm0.04$\\
 &  &   &  & Log--normal &  & $14.16\pm0.03$\\
 &  &   & Blue shear & One sided &  & $14.18\pm0.05$\\
 &  &   &  & Log--normal &  & $14.15\pm0.03$\\
\cline{3-7}
 &  & Cosmology  & Fiducial & One sided & $14.247\pm0.004$ & $14.17\pm0.04$\\
 &  &   &  & Log--normal &  & $14.15\pm0.03$\\
 &  &   & Blue shear & One sided &  & $14.15\pm0.05$\\
 &  &   &  & Log--normal &  & $14.13\pm0.03$\\
\enddata
\tablecomments{
GGL derived masses for bins of eROSITA clusters.
The bins are defined in terms of the reported redshift and halo mass in the eRASS1 catalogs.
For the $0.1<z<0.2$ samples, we consider both the primary eRASS1 catalog and the higher purity ``cosmology" sample.
We report the halo masses inferred from fiducial DES~Y3 shear and DES~Y3 blue shear.
We also consider variations of the GGL fitting, including selecting FLAMINGO halos by a one sided selection (all central halos above a given mass) or drawing halos from a log-normal distribution at a given central mass.
For brevity, we only report the results using the fiducial FLAMINGO simulation.
}
\end{deluxetable*}

\bibliography{paper}%

\end{document}